\documentclass[a4paper,fleqn,usenatbib]{mnras}

\usepackage{mathptmx}
\usepackage[T1]{fontenc}
\usepackage{ae,aecompl}
\usepackage{subfigure}

\usepackage{graphicx}	% Including figure files
\usepackage{amsmath}	% Advanced maths commands
\usepackage{amssymb}	% Extra maths symbols

\title[Star Formation Quenching in Green Valley Galaxies at $0.5\lesssim z\lesssim1.0$]{Star Formation Quenching in Green Valley Galaxies at $0.5\lesssim z\lesssim1.0$ and Constraints with Galaxy Morphologies}

\author[Nogueira-Cavalcante et al.]{J. P. Nogueira-Cavalcante,$^{1,2}$\thanks{E-mail: jpncavalcante@on.br} T. S. Gon\c calves,$^{1}$ K. Men\'endez-Delmestre,$^{1}$ \and and K. Sheth$^{3,4}$ \\
$^{1}$Observat\'orio do Valongo, Universidade Federal of Rio de Janeiro, Ladeira Pedro Ant\^onio, 43, Sa\'ude 20080-090, Rio de Janeiro, Brazil\\
$^{2}$Observat\'orio Nacional, Rua Gal. Jos\'e Cristino 77, S\~ao Crist\'ov\~ao 20921-400 Rio de Janeiro RJ, Brazil\\
$^{3}$National Radio Astronomy Observatory, 520 Edgemont Road, Charlottesville, VA 22903, USA\\
$^{4}$NASA HQ, 300 E Street SW, Washington DC 20546, USA\\
}

\date{Accepted XXX. Received YYY; in original form ZZZ}

\pubyear{2016}

\begin{document}
\label{firstpage}
\pagerange{\pageref{firstpage}--\pageref{lastpage}}
\maketitle

%%%%%%%%%%%%%%%%%%
%% 
%% Abstract of the paper 
%% 
%%%%%%%%%%%%%%%%%%

\begin{abstract}
We calculate the star formation quenching timescales in green valley galaxies at intermediate redshifts ($z\sim0.5-1$) using stacked zCOSMOS spectra of different galaxy morphological types: spheroidal, disk-like, irregular and merger, dividing disk-like galaxies further into unbarred, weakly-barred and strongly-barred, assuming a simple exponentially-decaying star formation history model and based on the H$_{\delta}$ absorption feature and the $4000$ \AA ~break. We find that different morphological types present different star formation quenching timescales, reinforcing the idea that the galaxy morphology is strongly correlated with the physical processes responsible for quenching star formation. Our quantification of the star formation quenching timescale indicates that disks have typical timescales $60\%$ to 5 times longer than that of galaxies presenting spheroidal, irregular or merger morphologies. Barred galaxies in particular present the slowest transition timescales through the green valley. This suggests that although secular evolution may ultimately lead to gas exhaustion in the host galaxy via bar-induced gas inflows that trigger star formation activity, secular agents are not major contributors in the rapid quenching of galaxies at these redshifts. Galaxy interaction, associated with the elliptical, irregular and merger morphologies contribute, to a more significant degree, to the fast transition through the green valley at these redshifts. In the light of previous works suggesting that both secular and merger processes are responsible for the star formation quenching at low redshifts, our results provide an explanation to the recent findings that star formation quenching happened at a faster pace at $z\sim0.8$.

\end{abstract}


%%%%%%%%%%%%%%%%%%%%%%%%%%%%%%%%%%%%%%%%%%%%%%%%%%

%%%%%%%%%%%%%%%%% BODY OF PAPER %%%%%%%%%%%%%%%%%%

%%%%%%%%%%%%%%%%%%
%% 
%% Introduction
%% 
%%%%%%%%%%%%%%%%%%

\section{Introduction}\label{introduction}

The distribution of galaxies in the colour-magnitude diagram (CMD) displays two groups of galaxies: red galaxies occupy the region known as the red sequence, while blue galaxies populate the blue cloud \citep[e.g.,][]{Strateva2001, Baldry2004, Willmer2006}. In general, blue galaxies are actively star-forming systems that are rich in gas, while red galaxies are gas-poor and have little or no star formation. Between the red sequence and the blue cloud on the CMD there is a sparsely populated region with a smaller number density of galaxies. This region is known as the green valley \citep[e.g.,][]{Martin2007, Wyder2007, Salim2007}. 

One of the main physical differences between the blue cloud and the red sequence galaxies relies on their stellar population content: red galaxies have, on average, older stellar populations whereas the stellar populations in blue galaxies are predominantly young \citep[e.g., ][]{Kauffmann2003}. Moreover, \citet{Pan2013} have shown that stellar populations in green valley galaxies are older than those in blue cloud galaxies and younger than those in red sequence galaxies. This supports an evolutionary scenario where blue galaxies evolve into red ones, transitioning through the green valley. Furthermore, the galaxy processes responsible for this transition must act rapidly to explain the scarcity of galaxies in this region of the CMD. 

Many astrophysical processes are well recognized to be agents that can either trigger star-formation (e.g., mergers, disk instabilities) $-$ thus leading to an accelerated exhaustion of the gas content in a galaxy $-$ or directly quench the star-forming activity in a galaxy by expelling or heating the necessary fuel for continued star formation (e.g., AGN feedback). Major mergers have been shown to lead to strong gas inflows that trigger galaxy-wide starbursts \citep{DiMatteo2005}; they can further lead to the growth of bulges and ultimately change the overall galaxy morphology \citep{Cheung2012, Springel2005d}. Feedback from supernovae in low-mass galaxies and active galactic nuclei (AGN) in massive galaxies have also been proposed as important quenching agents by either expelling or heating the gas that would otherwise collapse and form stars. In the case of massive galaxies, \citet{Nandra2007} studied a sample of X-rays sources and found that the majority of host galaxies are red, suggesting a scenario where the AGNs either cause or maintain the star formation quenching. Interestingly, \citet{Schawinski2009} found evidences of recent destruction of the molecular gas in a sample of faint active galactic nuclei, suggesting that even low-luminosity AGN episodes are able to quench star formation in galaxies. In the case of low-mass galaxies, supernovae winds have been shown to be capable of quenching star formation, expelling the gas from the interstellar medium \citep{Lagos2013, Menci2005}. Internal secular processes including disk instabilities (e.g., bars) have been associated with significant gas inflow responsible for the rapid consumption of gas through nuclear starbursts and the formation of pseudobulges \citep{Ho1997, Kormendy2004, Sheth2005}. Finally, external secular processes driven by the environment, such as strangulation, ram-pressure stripping and harassment, have been shown to remove or heat gas in galaxies as they enter high-density regions \citep{Coil2008, Dekel2003, Farouki1981, 
Gunn1972, Larson1980, Mendez2011, Moore1998}.

Although many of these processes have been the target of multiple studies within the context of star formation quenching, their relative importance as a function of cosmological time still remains an open question. At early times $(z \sim 1-2)$  \citet{Peng2010} suggest that fast processes (e.g., major mergers) play a very important role in the evolution of galaxies. However, \citet{Mendez2011}, using quantitative morphological parameters, found that the green valley at $0.4<z<1.2$ shows lower merger fractions than those measured for blue galaxies, suggesting that mergers are not important for quenching star formation in green valley galaxies at these redshifts. At later times ($z\sim0$), contrastingly, mergers become less common \citep{Conselice2003}, and slower processes that generally involve interactions between stars, gas clouds and the dark matter halo (e.g., disk instabilities, bars) may become more important \citep{Spitzer1951, Spitzer1953}. However, \citet{Martin2007} found that $\sim50\%$ of the green valley galaxies at $z\sim0.1$ show signs of AGN activity, although the AGN luminosity was not found to be correlated with the timescale of star formation quenching. Considering that AGN activity is usually associated with fast star formation quenching, as demonstrated by hydrodynamical simulations \citep{Dubois2013, Sijacki2006}, faster processes may also play an important role in quenching star formation at lower redshifts. These results show that we do not have a complete picture of the timescales and processes responsible for galaxy transition through the green valley either at local or distant universe.

Recent studies have shown that the pace of the galaxy transition from the blue cloud to the red sequence varies with cosmic time. With measurements of the star formation quenching timescales, stellar masses and galaxy number densities on the CMD, \citet{Martin2007} and \citet{Goncalves2012} determined the mass flux through the green valley to be $\sim$ 0.033 M$_{\odot}$ yr$^{-1}$ Mpc$^{-3}$ and 0.16 M$_{\odot}$ yr$^{-1}$ Mpc$^{-3}$ at $z\sim0.1$ and $z\sim0.8$, respectively. These studies showed that not only galaxies at intermediate redshifts transition through the green valley at a faster pace than those at low redshifts, but that more massive systems are transiting the green valley in the distant universe. With such a galaxy flow through the green valley, \citet{Goncalves2012} demonstrate a match to the buildup of the red sequence \citep{Faber2007} at a time prior to the establishment of the Hubble sequence observed today \citep{Delgado-Serrano2010}.

In this paper we investigate star formation quenching in the green valley as a function of galaxy morphology at these intermediate redshifts.  Different agents that trigger or quench star formation in galaxies leave morphological signatures in their host galaxies. This allows us to use galaxy morphology as a proxy for the astrophysical processes that may lead to star formation quenching. We determine star formation quenching timescales in galaxies that populate the green valley in the (NUV$-r$)~$\times$~M$_r$ colour-magnitude diagram at intermediate redshifts, in an effort to elucidate the role of different physical mechanisms on the green valley transition at a time coincident with a rapid growth of the red sequence. This study expands beyond the work of \citet{Goncalves2012} at $z\sim0.8$, who measured the star formation quenching timescales of green valley galaxies without distinguishing between different galaxy types or the physical mechanisms responsible for the quenching. We seek to understand the physical processes that are responsible for the mass flux through the green valley at $0.5\lesssim z\lesssim1.0$.

This paper is organized as follows. In Section \ref{methodology} we present the methodology used to measure the star formation quenching timescales based on galaxy spectra and colours. In Section \ref{galaxy_sample} we describe the galaxy data used in this work. In Section \ref{results} we show our results and in Section \ref{discussion} we discuss the possible galaxy mechanisms that might explain the star formation quenching in our galaxy data. Finally, in Section \ref{summary} we summarize our findings and present our current conclusions. We use standard cosmology throughout the paper, with $\Omega_\text{M}=0.3$, $\Omega_{\Lambda}=0.7$ and $h=0.7$. 

%%%%%%%%%%%%%%%%%%
%% 
%% Methodology
%% 
%%%%%%%%%%%%%%%%%%
% 
\section{Methodology}\label{methodology}

\begin{figure*}
\begin{center}
\includegraphics[width = 15cm]{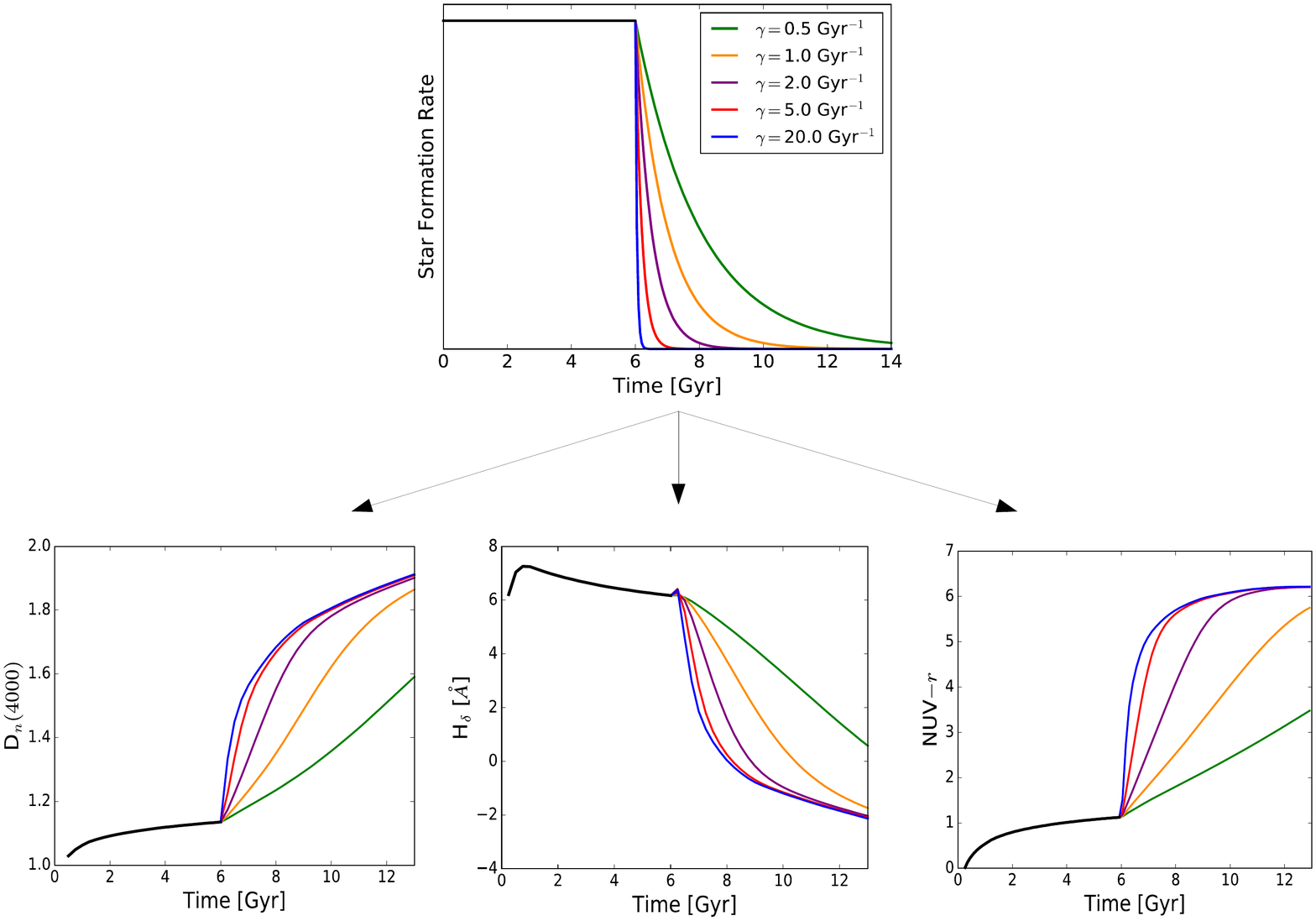}
\caption{\textit{\textbf{Top}}: Star formation rate as a function of time for five different SFHs with exponential decay (equations \ref{star_formation_history_1} and \ref{star_formation_history_2}). \textit{\textbf{Bottom}}: D$_n(4000)$ and  H$_{\delta,A}$ indices and NUV$-r$ colour as functions of time, assuming the exponentially-decaying SFHs with different $\gamma$ indices shown on the top panel. The faster the star formation quenching (i.e., higher $\gamma$ value) is, the sharper the time variation of these indices will be.}
\label{Dn_4000_H_delta_A_and_NUV_R_evolution}
\end{center}
\end{figure*}

To quantify star formation quenching timescales, we adopt the approach used by \citet{Martin2007} and \citet{Goncalves2012}. We make a simplifying assumption on the star formation history (SFH) for each galaxy, where the star formation rate is parametrized in the following way:	

 \begin{equation}
  \text{SFR}(t)=\text{SFR}(t=0), \,\,\,\,\ t<t_0 ;
  \label{star_formation_history_1}
 \end{equation}
 \begin{equation}
  \text{SFR}(t)=\text{SFR}(t=0)e^{-\gamma t}, \,\,\,\,\ t>t_0 \,\, ,
  \label{star_formation_history_2}
 \end{equation}
where $t_0$ is a characteristic time. Equations \ref{star_formation_history_1} and \ref{star_formation_history_2} tell us that the SFH for each galaxy is described by a constant star formation rate for the first $t_0=6$ Gyrs \citep{Goncalves2012} $-$ corresponding to a constant star formation rate up until redshift $z\sim0.8$ $-$ followed by a period of exponential decay. The $\gamma$ index, in units of Gyr$^{-1}$, characterizes the \textquotedblleft rate\textquotedblright \ of star formation quenching and is the quantity that we aim to measure. A smaller $\gamma$ value corresponds to slower quenching, whereas a larger one corresponds to faster quenching (see Figure \ref{Dn_4000_H_delta_A_and_NUV_R_evolution}).

The exponential index $\gamma$ can be obtained by applying the methodology described in \citet{Kauffmann2003}, based on measuring the rest-frame 4000 \AA \ break and the strength of the H$_{\delta}$ absorption line. The 4000 \AA \ break corresponds to a discontinuity in the optical spectrum of a galaxy caused mainly by the opacity in the metal-rich stellar atmospheres: the 4000 \AA \ break will be small in a galaxy dominated by young stellar populations because the metals in the atmospheres of O and B stars that dominate the galaxy's optical spectrum are multiply ionized and the ultraviolet radiation is not absorbed. The 4000 \AA \ break becomes larger with older stellar populations. We quantify the 4000 \AA \ break with the index defined in \citet{Balogh1999}:
\begin{equation}
  \text{D}_n(4000) = \sum_{\lambda = 4000 \text{\AA}}^{4100 \text{\AA}} F_{\lambda} \left/ \sum_{\lambda=3850 \text{\AA}}^{3950\text{\AA}} F_{\lambda} \right. \,\, , 
 \label{Dn_4000_indice}
\end{equation} 
where $F_{\lambda}$ is the flux at wavelength $\lambda$. The H$_{\delta}$ absorption is strongest for galaxies that experienced a star formation burst in $0.1-1$~Gyr ago, after O and B stars have finished their evolution, when the stellar light is dominated by A stars. H$_{\delta,A}$ index follows the same definition as \citet{Worthey1997}:
\begin{equation}
 \text{H}_{\delta,A} = \sum_{\lambda=4083,5 \text{\AA}}^{4122,25 \text{\AA}}\left(1 - \frac{F_{\lambda}}{F_{\lambda,\text{cont}}}\right) d\lambda \,\, ,
 \label{H_delta_A_indice}
\end{equation}
where $F_{\lambda,\text{cont}}$ is the continuum flux. We define the continuum flux by fitting a straight line through the average flux density between 4041.60~\AA \ and 4079.75~\AA, bluewards of the H$_{\delta}$ absorption feature, and 4128.50~\AA \ and 4161.00~\AA, redwards. We note that, although previous works have used indices as a function of $F_\nu$ \citep[e.g.,][]{Bruzual1983,Balogh1999}, here we quote values as a function of $F_\lambda$. Observational and synthetic spectra are all presented in those units, and we have performed all index measurements. This ensures consistency between zCOSMOS data and population models, as well as an agreement with previous work by \citet{Martin2007} and \citet{Goncalves2012}.

\begin{figure*}
\begin{center}
\includegraphics[width = 8cm]{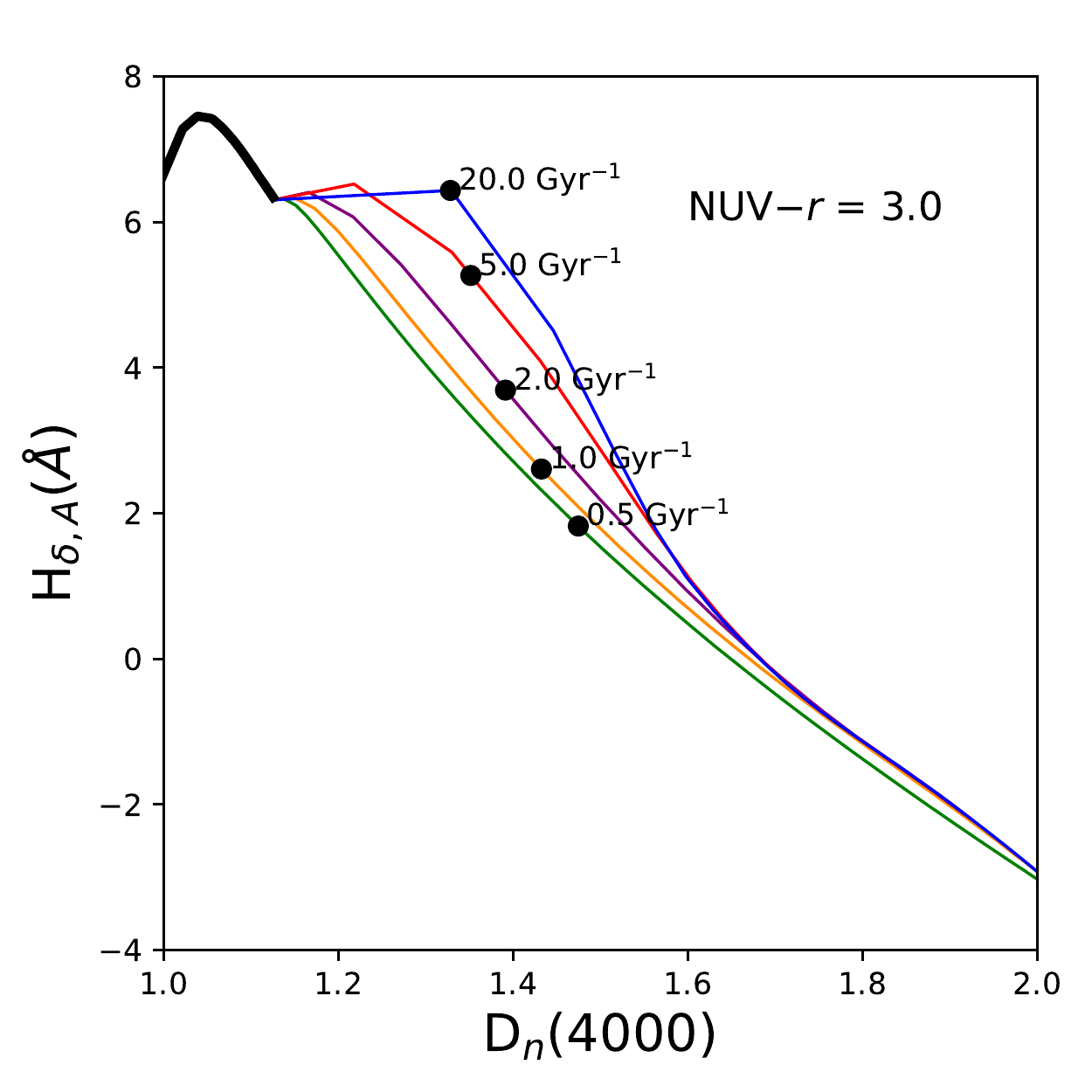}
\includegraphics[width = 8cm]{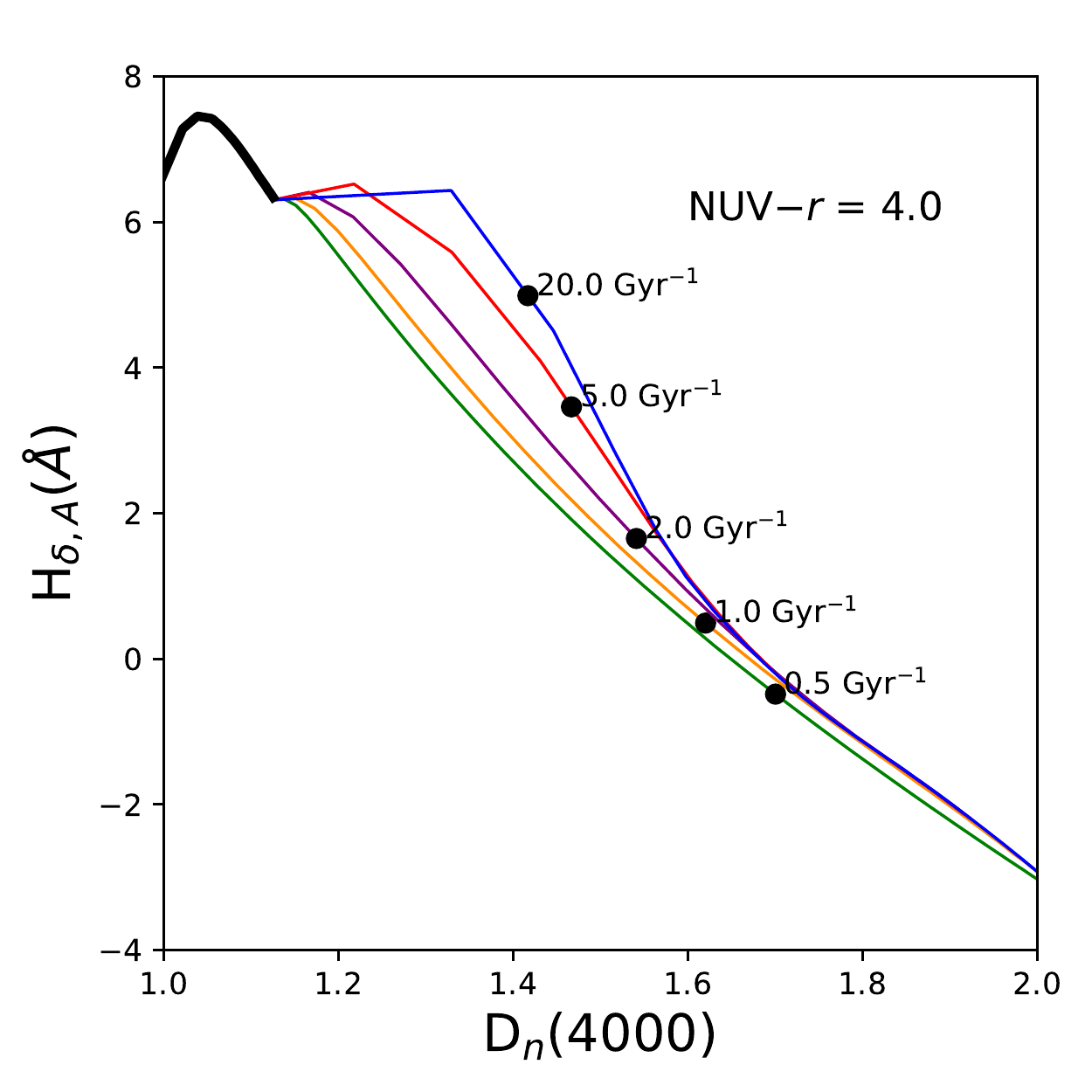}
\caption{H$_{\delta,A}$ $\times$ D$_n(4000)$ planes for the five SFH models shown in Figure \ref{Dn_4000_H_delta_A_and_NUV_R_evolution}. The black dots represent the H$_{\delta,A}$ and D$_n(4000)$ values for a given SFH model and NUV$-r$ colour. We can see clearly that different NUV$-r$ colours lead to different H$_{\delta,A}$ and D$_n(4000)$ values in the same SFH models.}
\label{H_delta_A_vs_Dn_4000_plane}
\end{center}
\end{figure*}

The H$_{\delta,A}$ and D$_n(4000)$ indices are sensitive to the galaxy SFH \citep{Kauffmann2003}. Both indices, as well as optical colours, are observables that depend strongly on the SFH of each galaxy. Figure \ref{Dn_4000_H_delta_A_and_NUV_R_evolution} shows how the H$_{\delta,A}$ and D$_n(4000)$ spectral indices, as well as the NUV$-r$ colour, change with time for five different examples of SFHs, with different star formation quenching indices ($\gamma$). We create a grid of curves in the H$_{\delta,A} \times$ D$_n(4000)$ plane for model galaxies $-$ based on synthetic spectra $-$ with different star formation quenching indices (see Figure \ref{H_delta_A_vs_Dn_4000_plane}). Along each curve, we identify the position of the NUV$-r$ colours that correspond to that of green valley galaxies. Figure \ref{H_delta_A_vs_Dn_4000_plane} shows how the H$_{\delta,A}$ and D$_n(4000)$ indices can assume different values for galaxies with the same SFH models but different colours. The galaxy models in this work were produced using the \citet{Bruzual2003} code, with \citet{Chabrier2003} initial mass functions, Padova 1994 stellar evolutionary tracks \citep{Alongi1993, Bressan1993, Fagotto1994a, Fagotto1994b} and solar metallicity.

\begin{figure*}

\begin{center}
\includegraphics[width = 7cm]{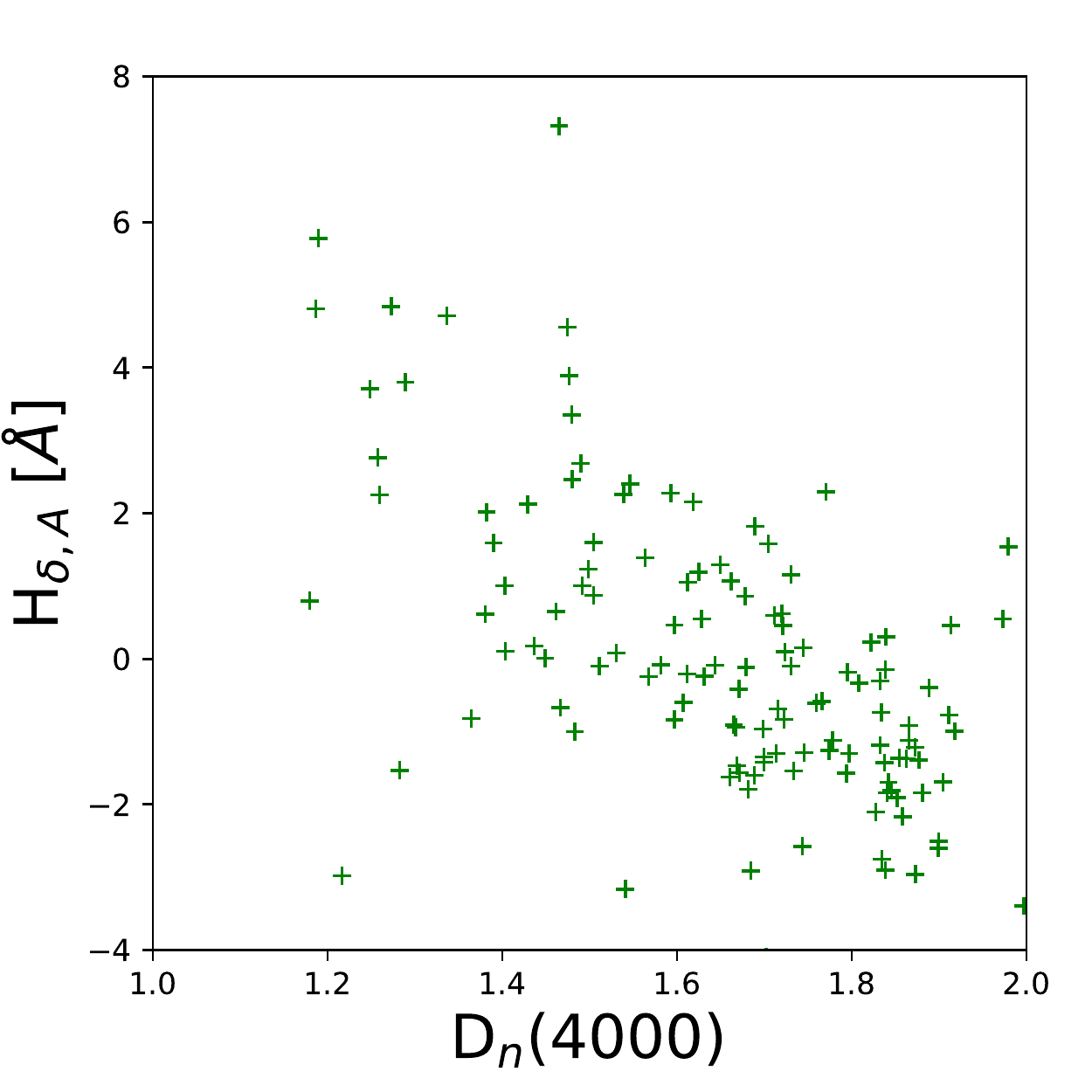}
\includegraphics[width = 7cm]{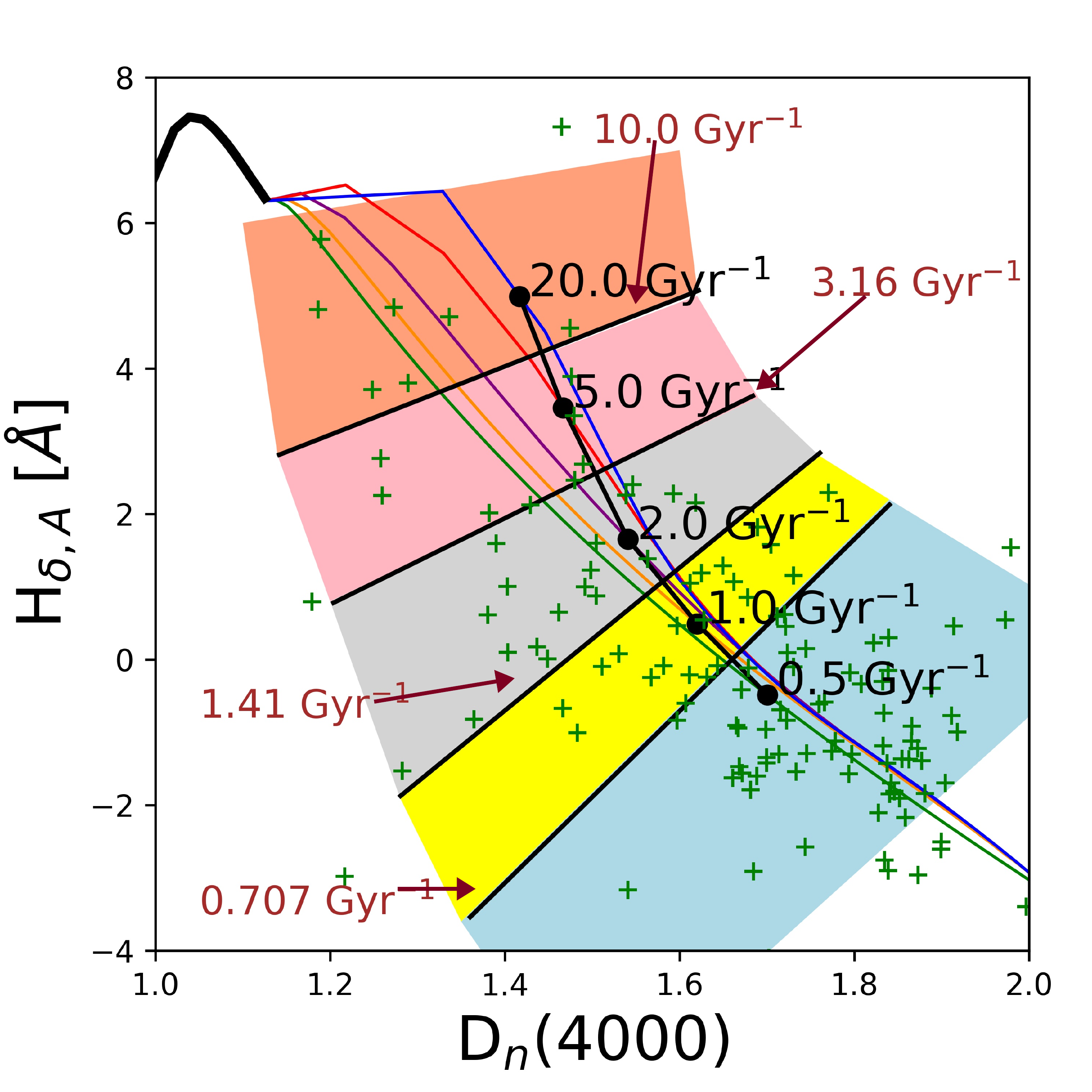}
\caption{\textit{\textbf{Left}}: Distribution of measured D$_n$(4000) and H$_{\delta,A}$ points from galaxy spectra of randomly SDSS green valley galaxies \citep{Martin2007}. \textit{\textbf{Right}}: Same diagram with the five SFH models considered in this work (see Figures \ref{Dn_4000_H_delta_A_and_NUV_R_evolution} and \ref{H_delta_A_vs_Dn_4000_plane}). Straight black lines represent the loci of geometric averages of $\gamma$ between two consecutive SFH models, as indicated by brown arrows. These are used to interpolate for the entire D$_n$(4000) -- H$_{\delta,A}$ plane, for each measurement of NUV$-r$ (see text for details). For this specific plot we consider NUV$-r=4.0$.}
%These straight lines delimit the dominant regions for each SFH model (light-blue for $\gamma = 0.5$ Gyr$^{-1}$, yellow for $\gamma = 1.0$ Gyr$^{-1}$, light-grey for $\gamma = 2.0$ Gyr$^{-1}$, light-pink for $\gamma = 5.0$ Gyr$^{-1}$ and salmon for $\gamma = 20.0$ Gyr$^{-1}$). 
\label{H_delta_A_vs_Dn4000_for_some_SDSS_GV_galaxies_with_SFH_models}
\end{center}
\end{figure*}

In order to take into account uncertainties in the exact SFH of real galaxies, we follow the same methodology to extract $\gamma$ indices from our measurements as \citet{Martin2007} and \citet{Goncalves2012}. More specifically, using consecutive pairs of points corresponding to the NUV$-r$ colour of the measured object, we define four perpendicular bisectors that represent the geometric loci where $\gamma$ equals the geometric mean of the two values represented by the points. We then interpolate the entire D$_n$(4000) $\times$ H$_{\delta,A}$ plane, mapping each coordinate to a singular $\gamma$ value from a set of three observables: NUV$-r$, D$_n$(4000) $\times$, and H$_{\delta,A}$. We repeat the procedure for every NUV$-r$ colour of each morphological type used in this work. This is represented in the right panel of Figure \ref{H_delta_A_vs_Dn4000_for_some_SDSS_GV_galaxies_with_SFH_models}.

%%%%%%%%%%%%%%%%%%
%% 
%% Figure 0
%% 
%%%%%%%%%%%%%%%%%%

%%%%%%%%%%%%%%%%%%
%% 
%% Figure 1
%% 
%%%%%%%%%%%%%%%%%%

%%%%%%%%%%%%%%%%%%
%% 
%% Figure 2
%% 
%%%%%%%%%%%%%%%%%%

\citet{Goncalves2012} used the same methodology to measure the mass flux through the green valley $-$ the quantity of mass per unit of time and volume traversing the green valley from the blue cloud to the red sequence $-$ at $z\sim0.8$. They find an agreement between their results and the buildup of the red sequence \citep{Faber2007}. These results exemplify the reliability of the methodology and furthermore provide a comparison sample of quenching indices at redshifts similar to our study.

%%%%%%%%%%%%%%%%%%
%% 
%% Sample
%% 
%%%%%%%%%%%%%%%%%%

\section{Galaxy Sample}\label{galaxy_sample}

We measure star formation quenching timescales in the green valley at intermediate redshifts ($0.5\lesssim z \lesssim1.0$) based on a sample of galaxies from the COSMOS\footnote{http://irsa.ipac.caltech.edu/data/COSMOS/} field \citep{Scoville2007}. We base our colour selection on photometry from the Canada-France-Hawaii Telescope Legacy Survey (CFHTLS), rely on the zCOSMOS spectroscopic survey (Data Release 2; \citealt{Lilly2007}) to obtain individual spectra for our green valley galaxies and use the morphological classification from the Zurich Estimator of Structure Types (ZEST) catalog \citep{Scarlata2007}. We detail each of these resources in this section.  

\subsection{Canada-France-Hawaii Telescope Legacy Survey}\label{canada_france_hawaii_telescope_legacy_survey}

We determine galaxy colours using photometry from the Canada-France-Hawaii Telescope Legacy Survey (CFHTLS) to define our green valley sample, following the same procedure as in \citet{Goncalves2012}. We select galaxies with detections in all five CFHTLS bands (u,g,r,i,z) brighter than m$_r < 24$, a necessary condition to correctly calculate the NUV and $r$ rest-frame magnitudes to properly separate the galaxy populations on the colour-magnitude diagram for galaxies at our redshifts of interest. Considering that red galaxies are very faint in the rest frame NUV (traced by the u and g bands at $0.5\lesssim z\lesssim1.0$), this approach disfavours the selection of red sequence objects. However, this does not significantly affect the completeness of green valley or blue cloud galaxies as much since, by definition, these objects are more luminous in the rest frame NUV. Moreover, the quenching timescales measured for green valley galaxies are not sensitive to the magnitude limit of the sample \citep[see Tables 3 and 4 in ][]{Goncalves2012}. The rest-frame magnitudes were calculated using the {\it K-correct} code \citep[version 4.2;][]{Blanton2007}. We also estimate uncertainties in our NUV$-r$ measurements. Typical observed photometric uncertainties in CFHTLS data are 0.1 mag or below. This translates to errors of approximately 0.15 mag in NUV$-r$ colours after k-correction is performed.

An important issue that must be taken into account in determining intrinsic galaxy colours is the contamination caused by dust, particularly in star-forming galaxies. We use the extinction correction model from \citet{Salim2009}: they compare stellar population models attenuated by dust with observed spectral energy distributions to derive galaxy properties, including the star formation rate. They find the following relation:
\begin{equation}
A_{\text{FUV}} = 3.68(\text{FUV}-\text{NUV})+0.29 \,\, ,
\end{equation}
where NUV and FUV are the rest-frame absolute magnitudes. To determine the extinction correction in other filters they use
\begin{equation}
 A_{\lambda}\propto \lambda^{-0.7} \,\, .
\end{equation}

\begin{figure}
\begin{center}

\includegraphics[width=8cm]{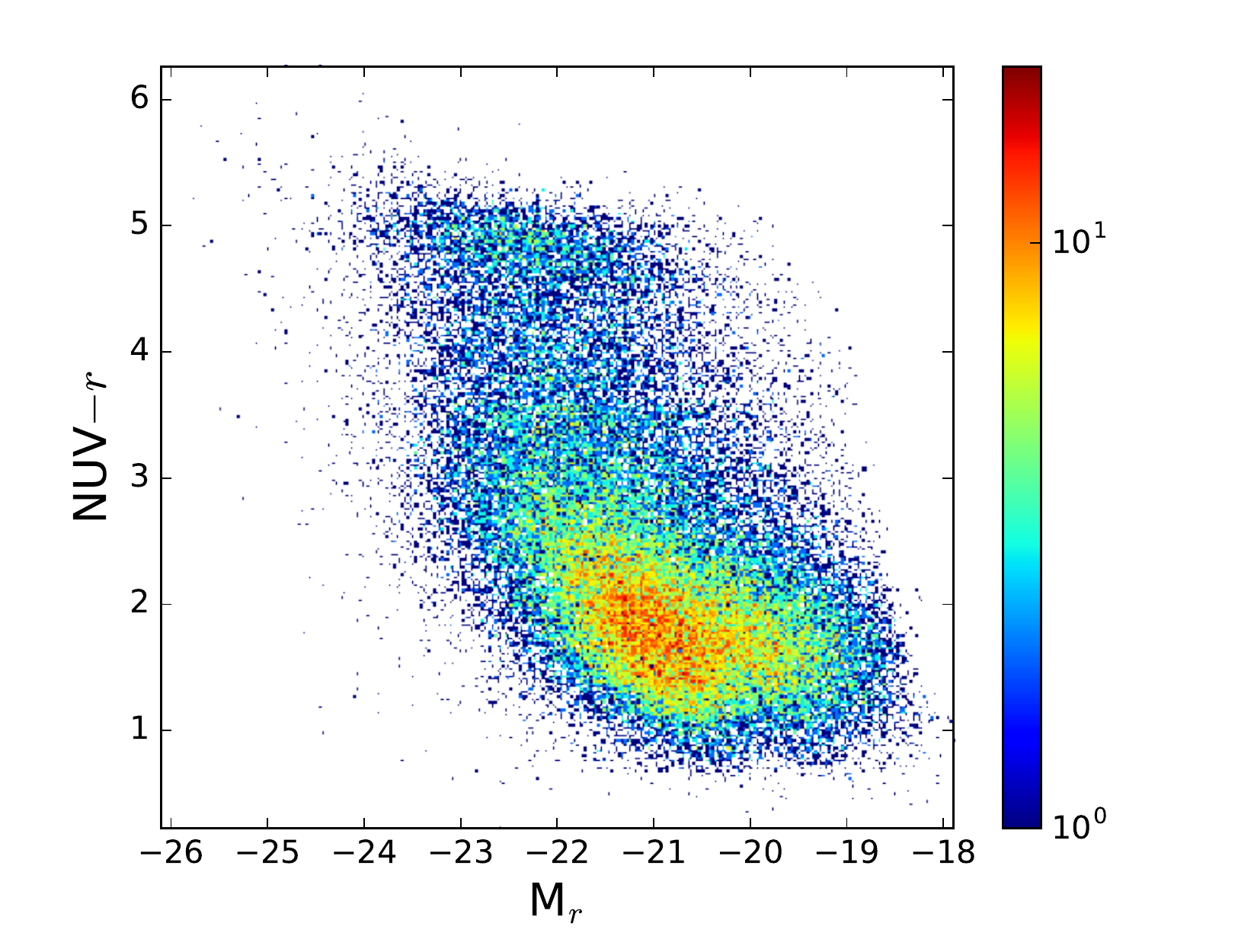}
\vspace{0.5cm}
\includegraphics[width=8cm]{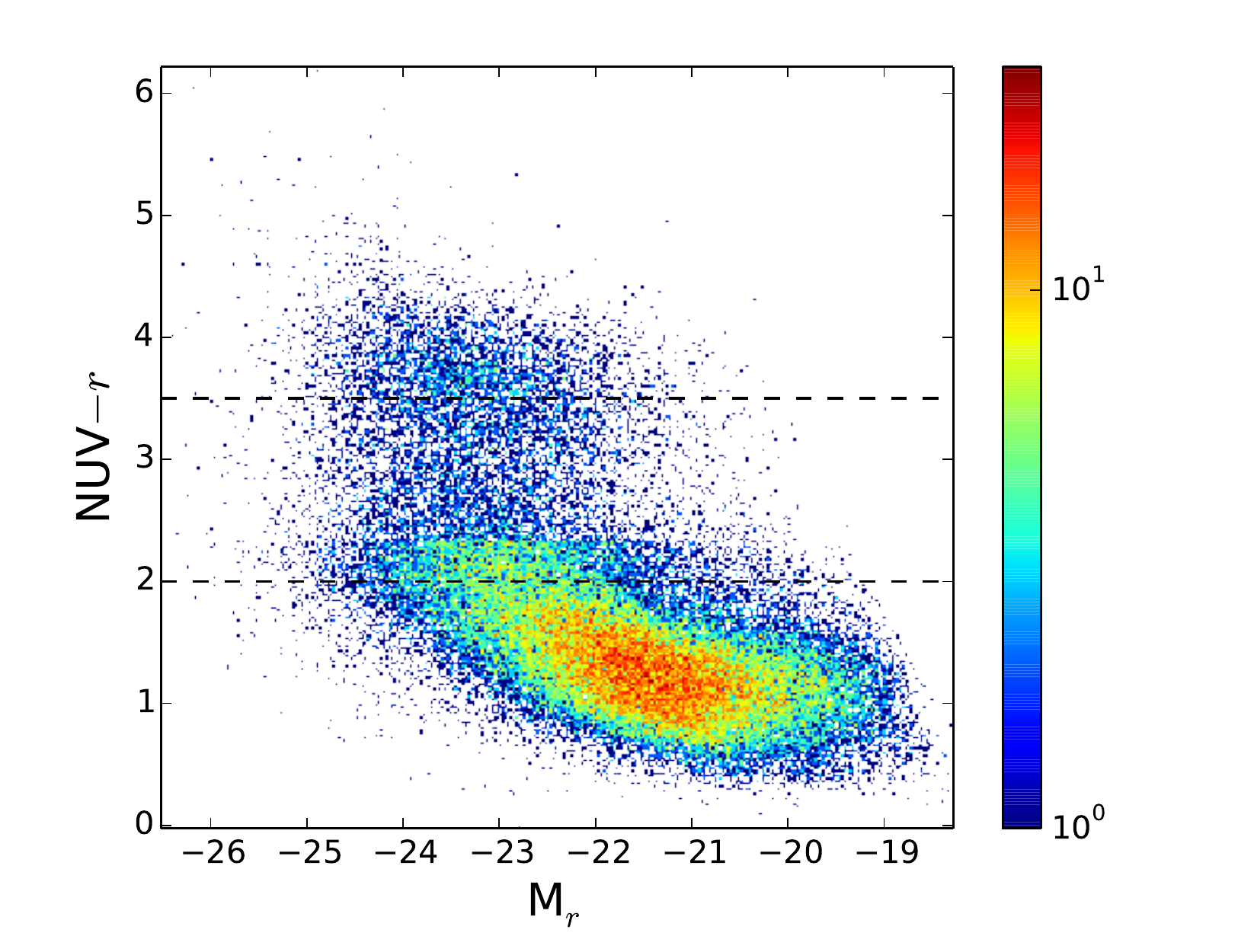}
\caption{Colour-magnitude diagrams of the Canada-France-Hawaii Telescope Legacy Survey galaxies used in this work. The top panel does not take into account the dust effects whereas the bottom panel is corrected by dust extinction. The horizontal axis represents the Sloan Digital Sky Survey $r$-band absolute magnitude and the vertical axis represents the colour NUV$-r$ which can best distinguish the galaxy populations (blue cloud, green valley and red sequence). The dashed horizontal lines delimit the green valley region. The red sequence is not well represented because a requirement of this work is that all galaxies must have all five CFHTLS photometric bands and the red galaxies are faint in the bluer bands.}
\label{colour_magnitude_diagram_CFHTLS}
\end{center}
\end{figure}

Figure \ref{colour_magnitude_diagram_CFHTLS} shows the colour-magnitude diagram of CFHTLS galaxies before (top) and after (bottom) correcting for dust attenuation. The procedure works well for star-forming galaxies, but overcorrects the colours of galaxies hosting older stellar populations \citep{Salim2009}. Stellar population synthesis models are not able to reproduce the (over-)corrected colours of galaxies in the green valley and the red sequence, and therefore we use this procedure only to remove contaminant dusty star-forming galaxies from the green valley region. We define our green valley galaxy sample by selecting objects with $2.0<$ NUV$-r<3.5$ in the dust-corrected CMD, but assume that the uncorrected rest-frame colours of this sample are effectively unaltered by dust for the remainder of the analysis.

%%%%%%%%%%%%%%%%%%
%% 
%% Figure 3
%% 
%%%%%%%%%%%%%%%%%%

\subsection{The 10k zCOSMOS Spectroscopic Sample}\label{the_zCOSMOS_spectroscopic_sample}

We exploit the zCOSMOS spectroscopic survey (Data Release 2; \citealt{Lilly2007}) to obtain spectra for our green valley galaxies and measure the D$_n(4000)$ and H$_{\delta,A}$ spectral indices. This survey comprises $\sim$10,000 spectra of galaxies in the COSMOS field with magnitudes down to $i=22.5$, taken with the VIsible Multi-Object Spectrograph (VIMOS) instrument located on the the ESO Very Large Telescope. The spectral coverage of these spectra is from 5500\AA \ to 9700\AA, with a resolution of $R \sim 600$.

We consider the possibility that the shallower magnitude limit of the spectroscopic sample might affect our results, since we are effectively averaging over measurements of more luminous galaxies by coadding these spectra. Nevertheless, as discussed above, \citet{Goncalves2012} have shown that there is no clear trend of quenching timescales as a function of luminosity. Still, in the future we intend to compare our results with deeper spectroscopic surveys, in order to verify whether there is a trend of timescales as a function of luminosity for any given morphological type.

\subsection{Galaxy Morphology Classification} \label{galaxy_morphology_classification}

The Zurich Estimator of Structure Types (ZEST) catalog \citep{Scarlata2007} classifies galaxies down to $i=24$ in the COSMOS field according to morphology based on a principal component analysis (PCA) of the galaxy structure. We base ourselves on this catalog to separate galaxies in our sample into the following morphological classes: elliptical galaxies (T$_{ZEST}$ = 1), disk galaxies (T$_{ZEST}$ = 2) and irregular galaxies (T$_{ZEST}$ = 3). We note that the distribution of ZEST morphological classifications within all green valley galaxies down to $i=22.5$ (i.e., within the zCOSMOS spectroscopic sample) is dominated by ellipticals ($\sim$51\%), with significant contributions from disks ($\sim$33\%) and irregular galaxies ($\sim$16\%).

\citet{Scarlata2007} tested the ZEST morphologies using visual classifications for a $z=0$ sample and found that the ZEST classifications agrees well with published morphologies. However, this classification has its limitations at higher redshifts due to decreased spatial resolution that turns the parametric analysis of structural details increasingly challenging. For this reason we carefully inspected HST optical images (F814W, $I$ band) for all green valley galaxies within the zCOSMOS sample and excluded from our sample the galaxies corresponding to the following cases: ZEST-classified ellipticals with evident tidal- or disk-like structures, ZEST-classified disks displaying multiple clumps reminiscent of coalesced mergers and ZEST-classified irregulars where a clear separation between galaxy pairs (or multiple companions) was evidence of an early-stage merger. Considering that we refrained from re-classifying these galaxies into a different morphological type and rather opted for discarding them from our sample, this resulted in a sample reduction of 56\%, 30\% and 40\% for the elliptical, disk and irregular galaxy classes, respectively. Although we reduced the galaxy sample analyzed by $\sim$50\%, with this approach we ensure a higher degree of purity within the morphological classes in our analysis. The resulting distribution of morphologies within the green valley sample corresponds to 40\%  ellipticals, 43\% disks and 17\% irregulars.

Within the disk galaxy classification, we also identify galaxies containing weak or strong bars, based on the classification by \citet{Sheth2008}. These authors identified 796 barred disks from a sample of $\sim2000$ galaxies. Their bar classification was obtained using two independent methods: by visual classification and through the ellipse-fitting technique, which is based on inspection of the ellipticity and position angle profiles that result from fitting elliptical isophotes to the 2D light distribution of disk galaxies \citep[e.g,][]{Menendez-Delmestre2007}. Our resulting classification for disk galaxies is hence that of strongly barred, weakly barred and unbarred galaxies. 

\begin{figure*}
\begin{center}
\subfigure
{
\includegraphics[width=2cm]{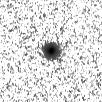}
\includegraphics[width=2cm]{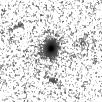}
\includegraphics[width=2cm]{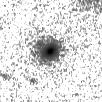}
\includegraphics[width=2cm]{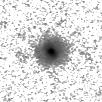}
\includegraphics[width=2cm]{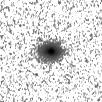}
\includegraphics[width=2cm]{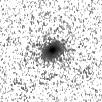}
\includegraphics[width=2cm]{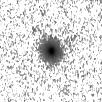}
\includegraphics[width=2cm]{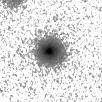}
}
\subfigure
{
\includegraphics[width=2cm]{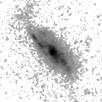}
\includegraphics[width=2cm]{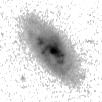}
\includegraphics[width=2cm]{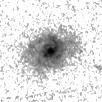}
\includegraphics[width=2cm]{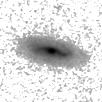}
\includegraphics[width=2cm]{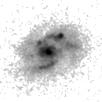}
\includegraphics[width=2cm]{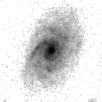}
\includegraphics[width=2cm]{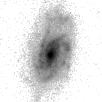}
\includegraphics[width=2cm]{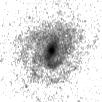}
}
\subfigure
{
\includegraphics[width=2cm]{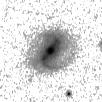}
\includegraphics[width=2cm]{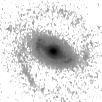}
\includegraphics[width=2cm]{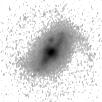}
\includegraphics[width=2cm]{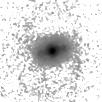}
\includegraphics[width=2cm]{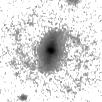}
\includegraphics[width=2cm]{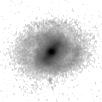}
\includegraphics[width=2cm]{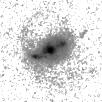}
\includegraphics[width=2cm]{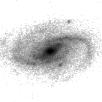}
}
\subfigure
{
\includegraphics[width=2cm]{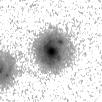}
\includegraphics[width=2cm]{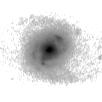}
\includegraphics[width=2cm]{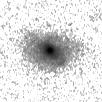}
\includegraphics[width=2cm]{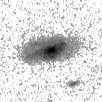}
\includegraphics[width=2cm]{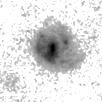}
\includegraphics[width=2cm]{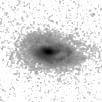}
\includegraphics[width=2cm]{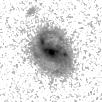}
\includegraphics[width=2cm]{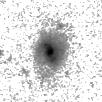}
}
\subfigure
{
\includegraphics[width=2cm]{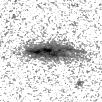}
\includegraphics[width=2cm]{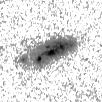}
\includegraphics[width=2cm]{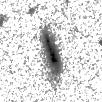}
\includegraphics[width=2cm]{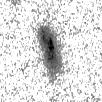}
\includegraphics[width=2cm]{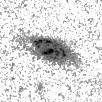}
\includegraphics[width=2cm]{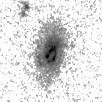}
\includegraphics[width=2cm]{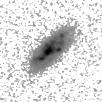}
\includegraphics[width=2cm]{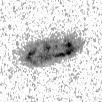}
}
\subfigure
{
\includegraphics[width=2cm]{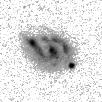}
\includegraphics[width=2cm]{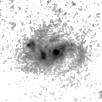}
\includegraphics[width=2cm]{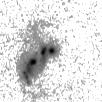}
\includegraphics[width=2cm]{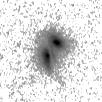}
\includegraphics[width=2cm]{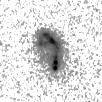}
\includegraphics[width=2cm]{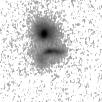}
\includegraphics[width=2cm]{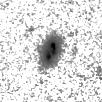}
\includegraphics[width=2cm]{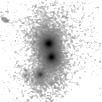}
}
\caption{Examples of green valley galaxy images in our sample taken with the Hubble Space Telescope (5'' $\times$ 5'' in size). From the top to the bottom rows these galaxies are classified as ellipticals, unbarred disks, strongly barred disks, weakly barred disks, irregulars and merging galaxies.}
\label{green_valley_galaxy_images}
\end{center}
\end{figure*}

For our analysis we also added the {\it merging} galaxy type, which comprises clear examples of (major) mergers. We identified these based on visual inspection, identifying the presence of tidal tails pointing to interactions between multiple galaxies. The number of galaxies for each galaxy type is shown in Table \ref{quenching_values}. The images of these galaxies are shown in Figure \ref{green_valley_galaxy_images}.

%%%%%%%%%%%%%%%%%%
%% 
%% Results
%% 
%%%%%%%%%%%%%%%%%%

\section{Results}\label{results}

We cross-matched the catalogs $-$ CFHTLS photometry, zCOSMOS spectra database, morphology classification from ZEST and \citet{Sheth2008} $-$ with an angular radius of 1'' and identified a total of 416 green valley galaxies with spectroscopic information from zCOSMOS, within the redshift interval of $0.5\lesssim z\lesssim1.0$. We have chosen this redshift interval in order to focus our study in understanding the physical processes responsible for the mass flux at $z\sim0.8$ found by \citet{Goncalves2012}. 

%%%%%%%%%%%%%%%%%%
%% 
%% Figure 4
%% 
%%%%%%%%%%%%%%%%%%

\begin{figure*}
\begin{center}
\includegraphics[width=5.5cm]{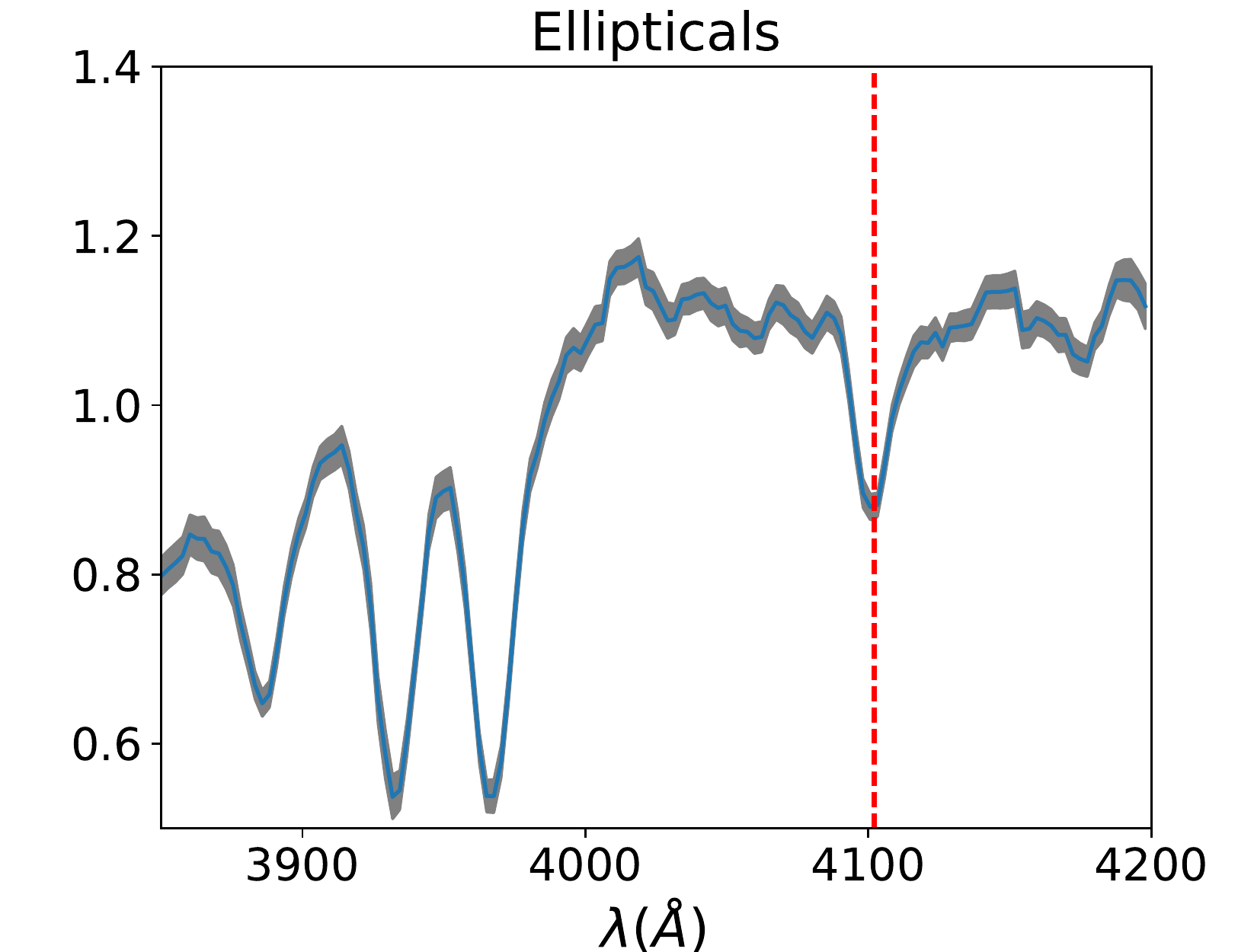}
\includegraphics[width=5.5cm]{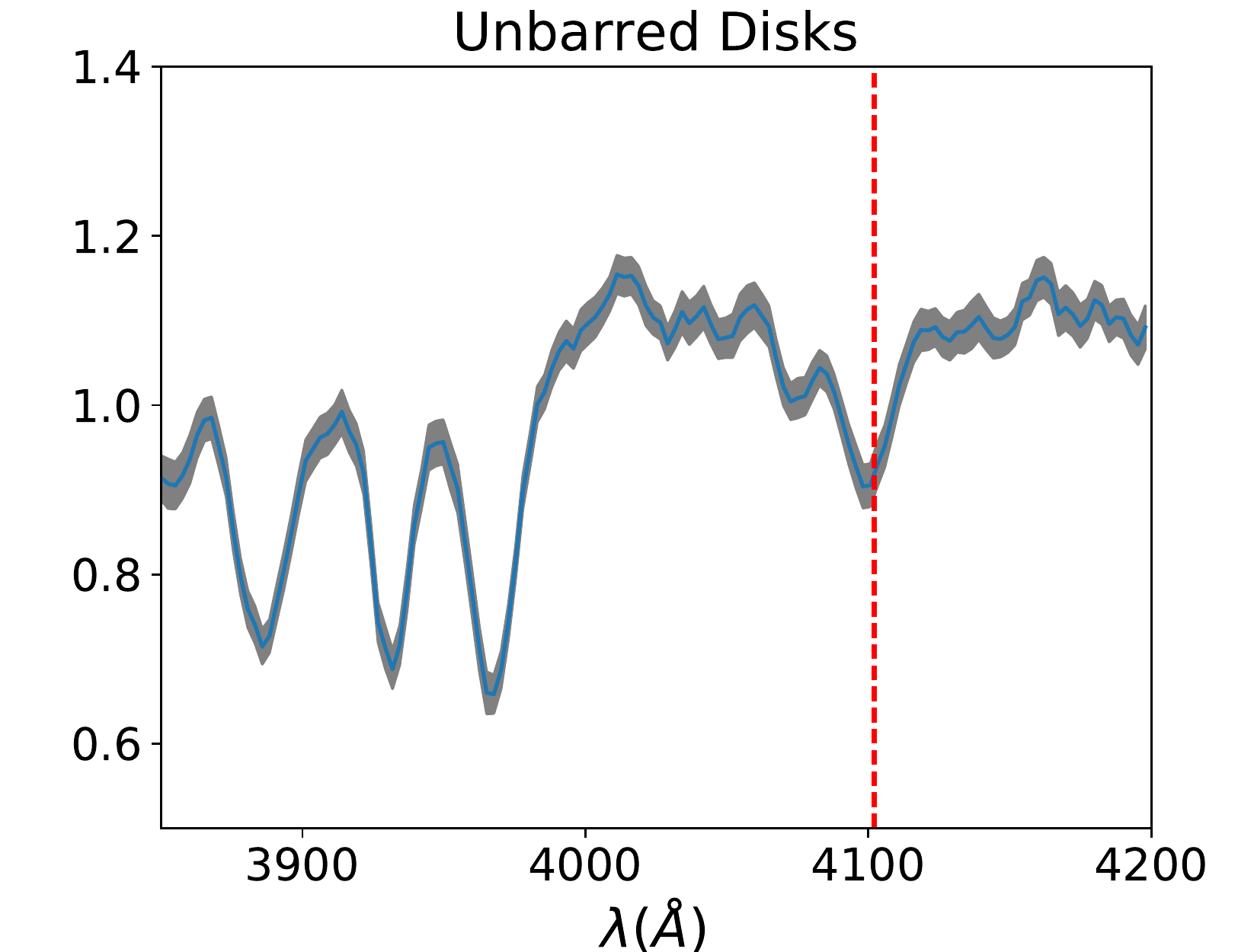}
\vspace*{0.5cm}
\includegraphics[width=5.5cm]{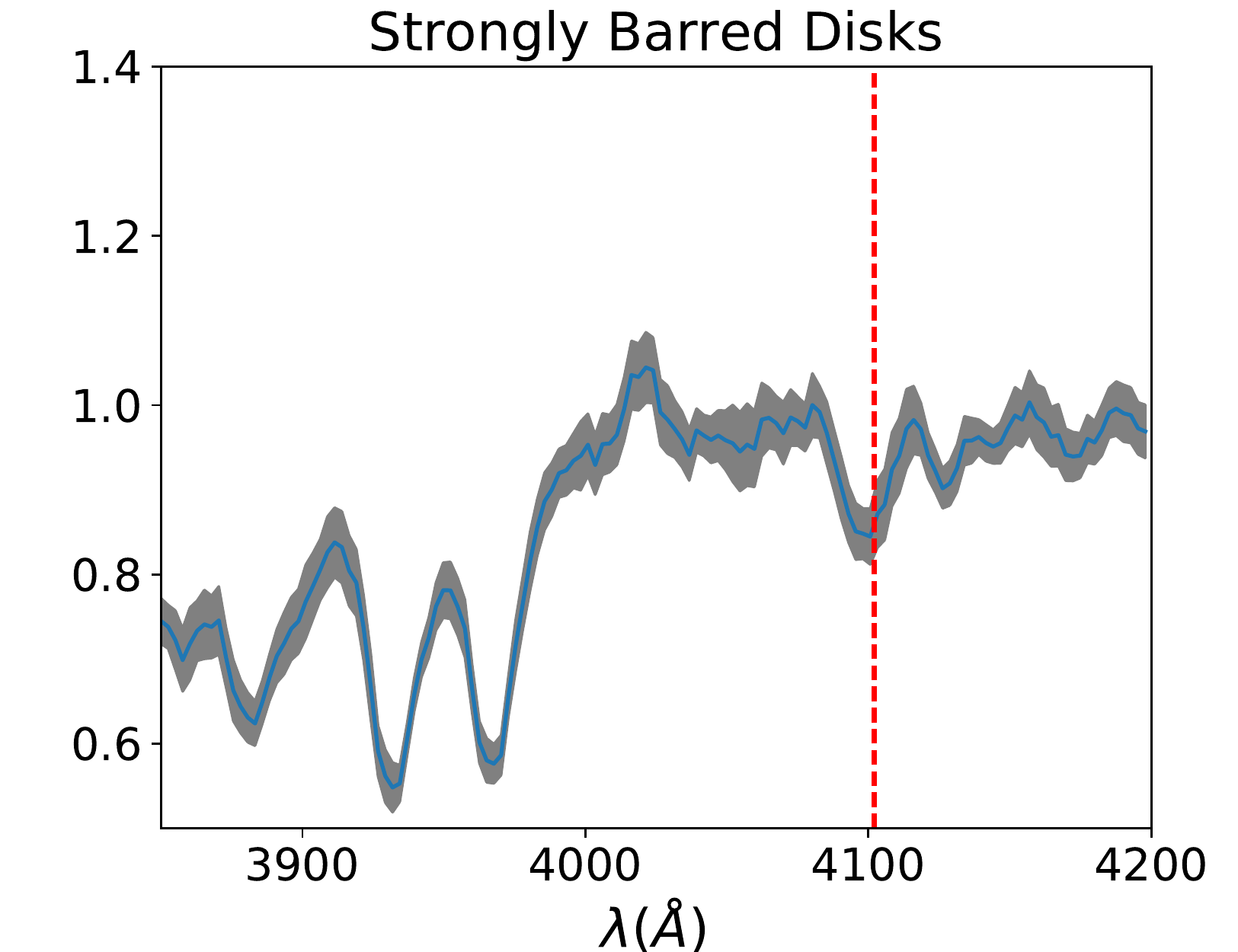}
\vspace*{0.5cm}
\includegraphics[width=5.5cm]{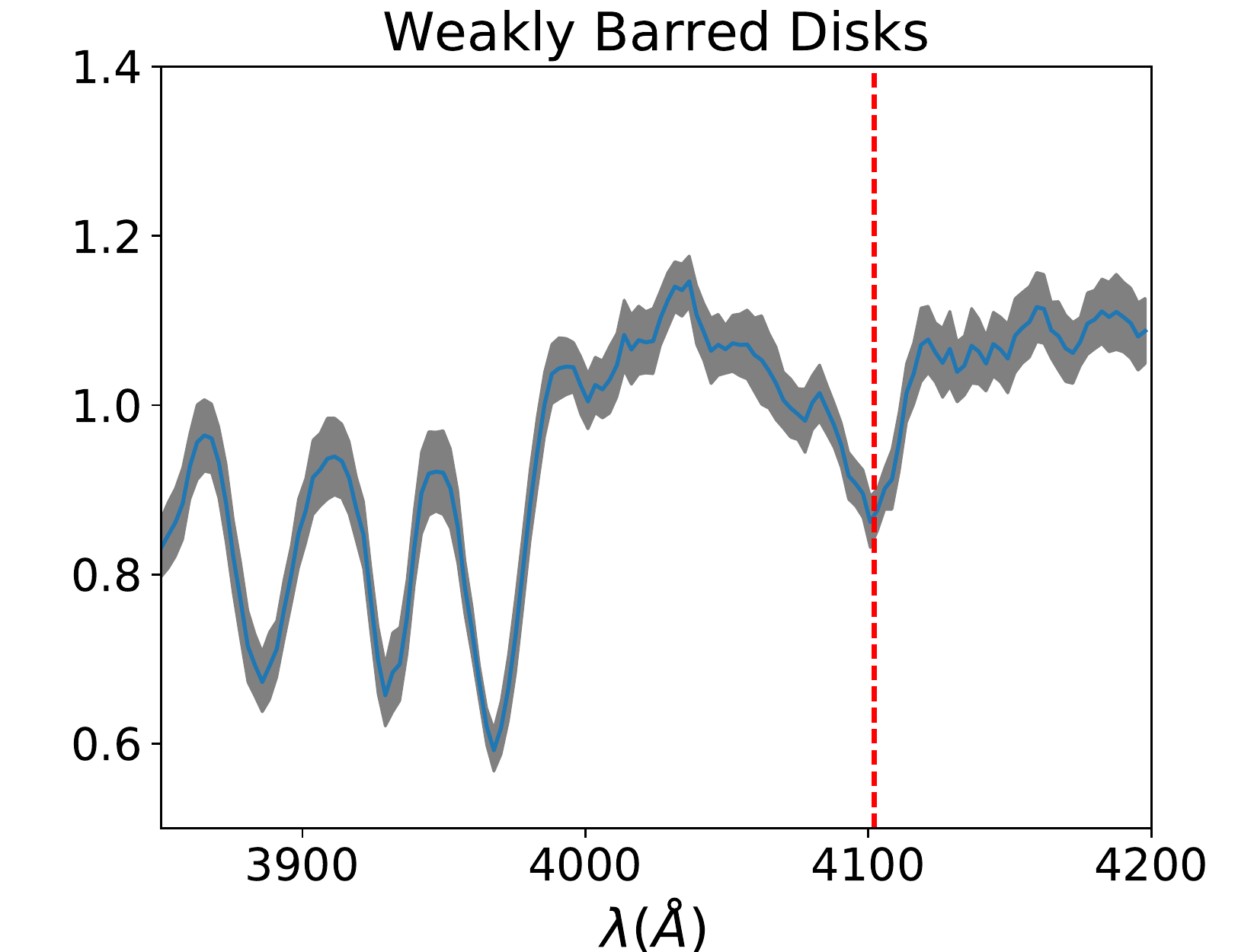}
\includegraphics[width=5.5cm]{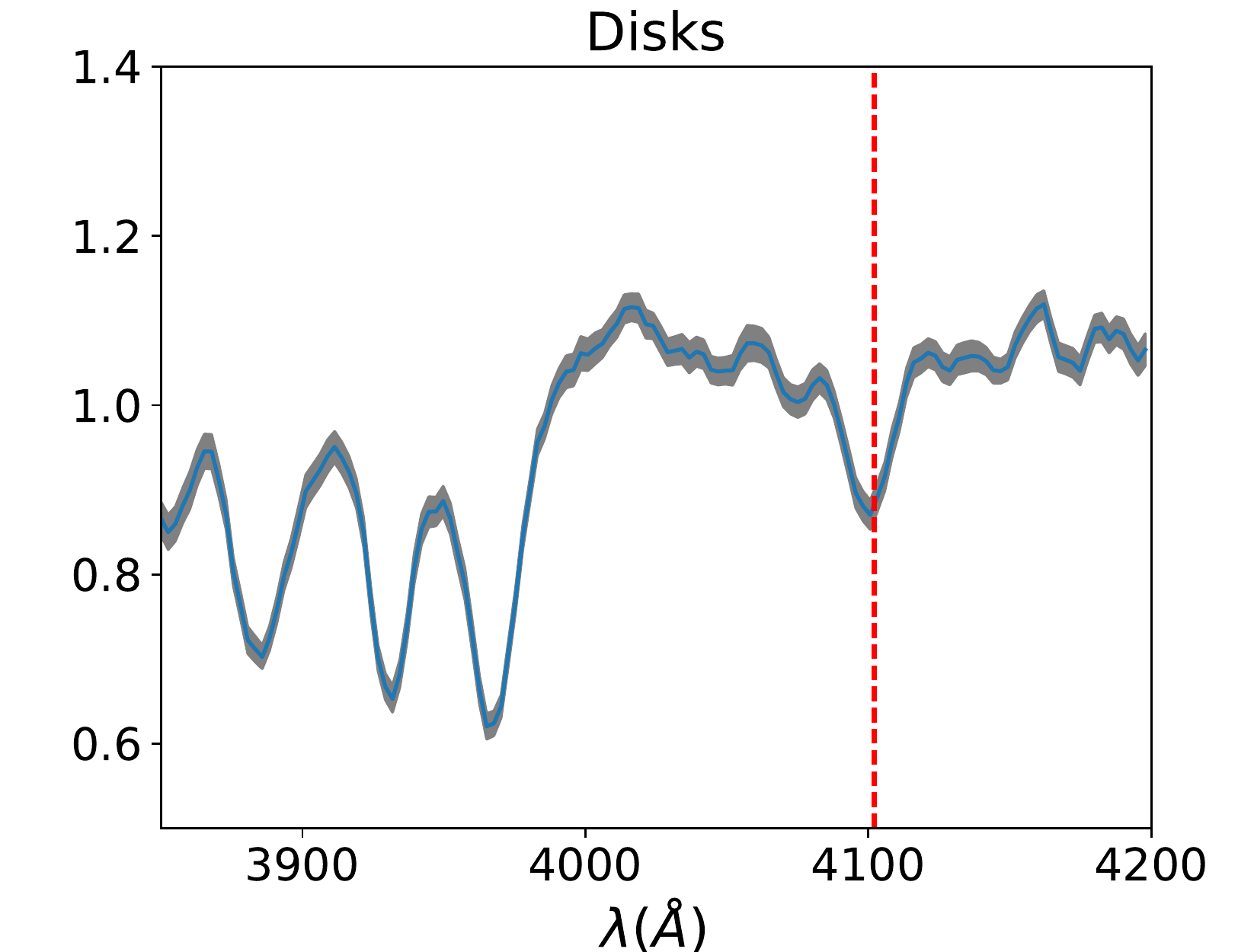}
\includegraphics[width=5.5cm]{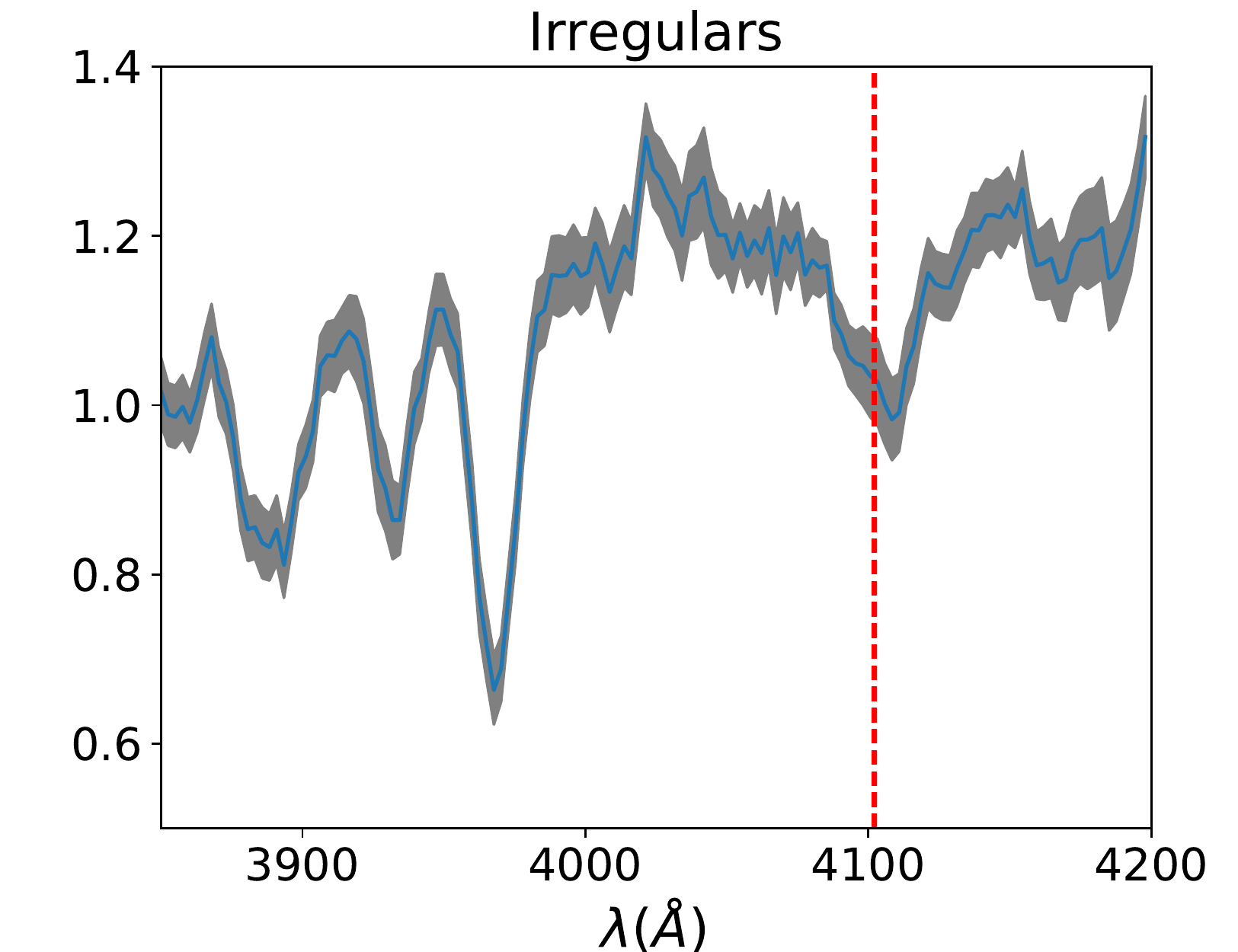}
\includegraphics[width=5.5cm]{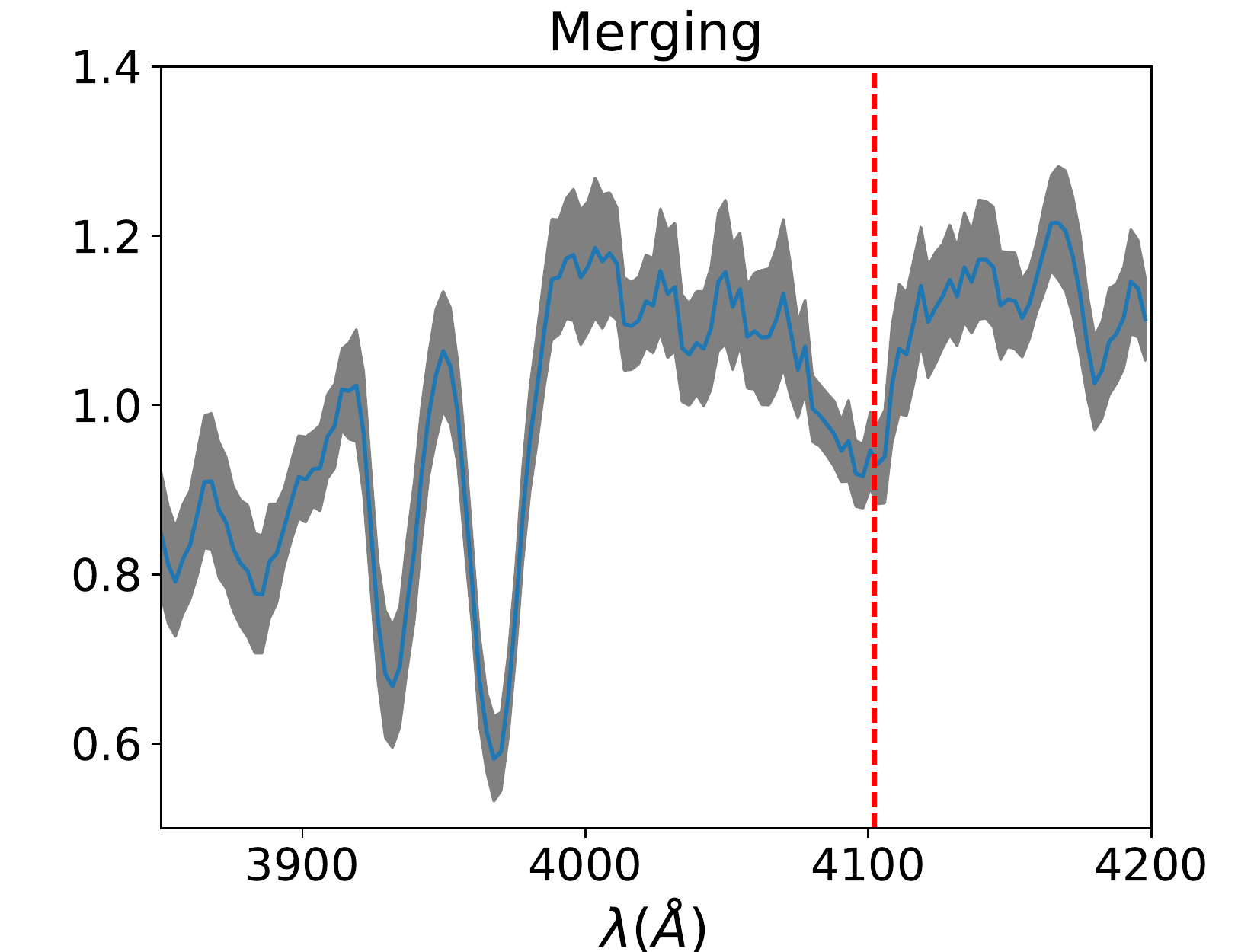}
\caption{Coadded spectra of the zCOSMOS galaxy samples according to morphological type. The grey region represents the standard deviation in each pixel. The red line highlights the H$_{\delta}$ absorption line.}
\label{coadded_zCOSMOS_spectra}
\end{center}
\end{figure*}

The signal-to-noise ratio (S/N) of individual zCOSMOS spectra is too low to calculate the D$_n$(4000) and H$_{\delta,A}$ spectral indices with reasonable uncertainties to distinguish between different SFHs. In order to increase the S/N we stacked the zCOSMOS spectra according to galaxy morphological type. We coadded the spectra calculating an initial median flux for each wavelength interval, identifying individual flux values beyond 2$\sigma$ and rejecting these to calculate a new median; the process is repeated until the rejected entries reach a given pre-determined percentage of the population, which we set to 2\% after confirming that it had no impact on the resulting coadded spectra. We created combined green valley galaxy spectra for disks, ellipticals, irregulars and merging galaxies by coadding all individual zCOSMOS spectra of galaxies corresponding to these morphological types. We also produced combined spectra for strongly barred, weakly barred and unbarred disk galaxies in order to extend our 
comparison to all disk subclasses. The resulting coadded spectra are shown in Figure \ref{coadded_zCOSMOS_spectra}.

%%%%%%%%%%%%%%%%%%
%% 
%% Figure 5
%% 
%%%%%%%%%%%%%%%%%%

We measure H$_{\delta,A}$ and D$_{n}$(4000) spectral indices for each combined spectrum in Figure \ref{coadded_zCOSMOS_spectra}. From these values, and considering the average NUV$-r$ colour for the set of galaxies that make up each morphological class, we are in principle in a position to calculate the empirical star formation quenching index, $\gamma$, using the SFH model discussed in Section \ref{methodology}. However, model spectra from \citet{Bruzual2003} have a higher spectral resolution (R$\sim3000$) than the zCOSMOS spectra (R$\sim600$). In order to make our results robust, we created the grid of H$_{\delta,A}$ vs. D$_n(4000)$ model curves for different star formation quenching indices based on \citet{Bruzual2003} spectra degraded down to the same resolution as zCOSMOS spectra.

\begin{figure*}
\begin{center}
\includegraphics[width=5.5cm]{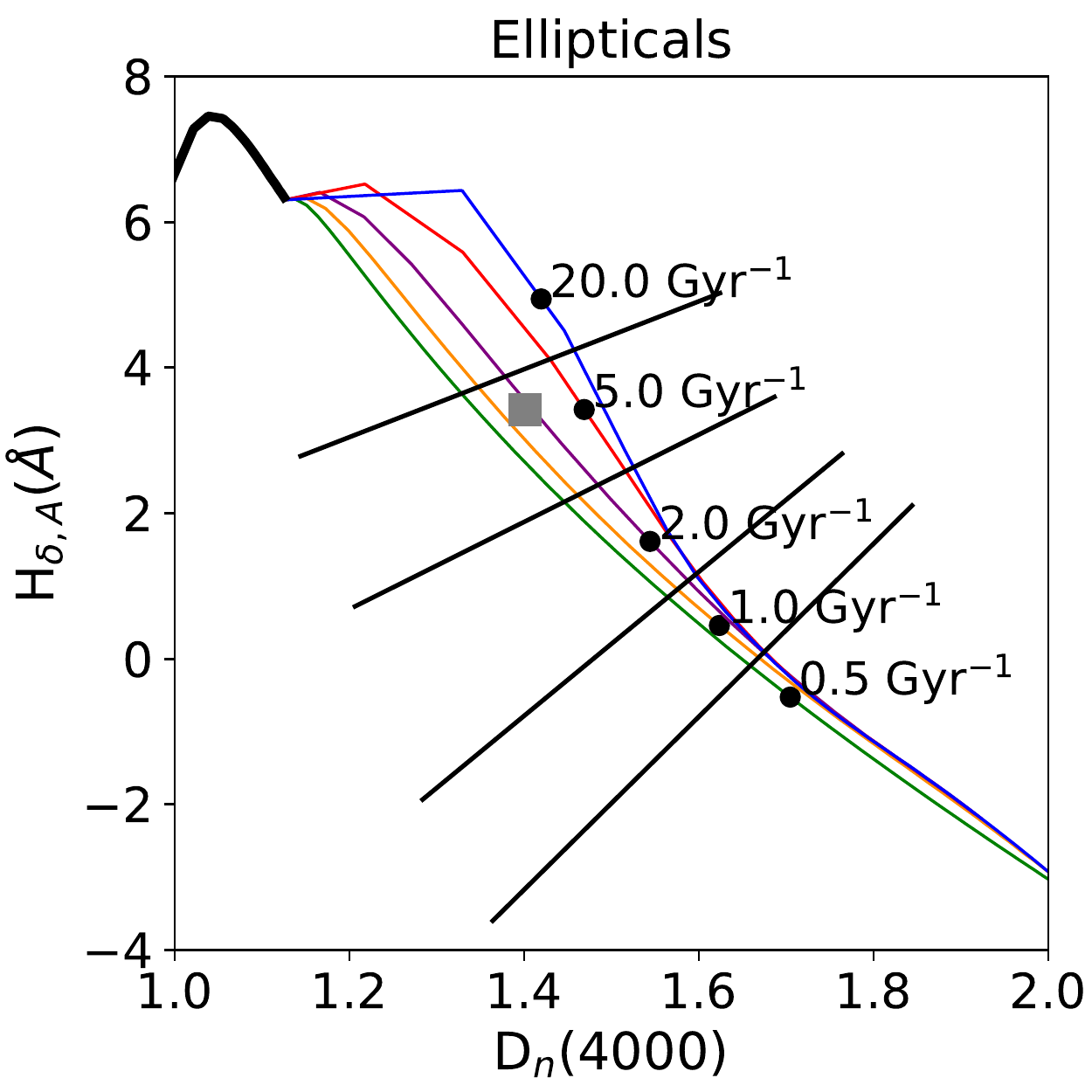}
\includegraphics[width=5.5cm]{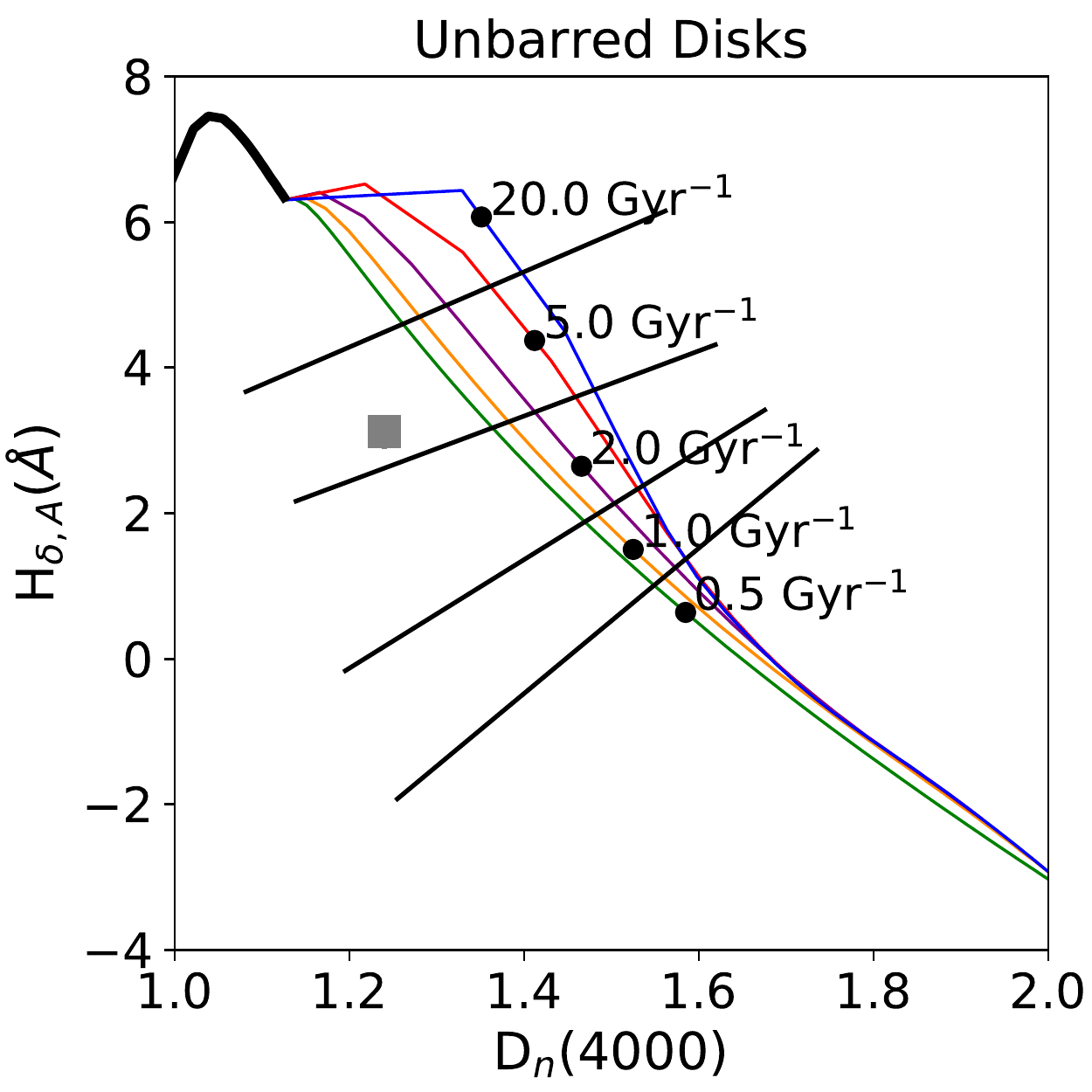}
\vspace*{0.5cm}
\includegraphics[width=5.5cm]{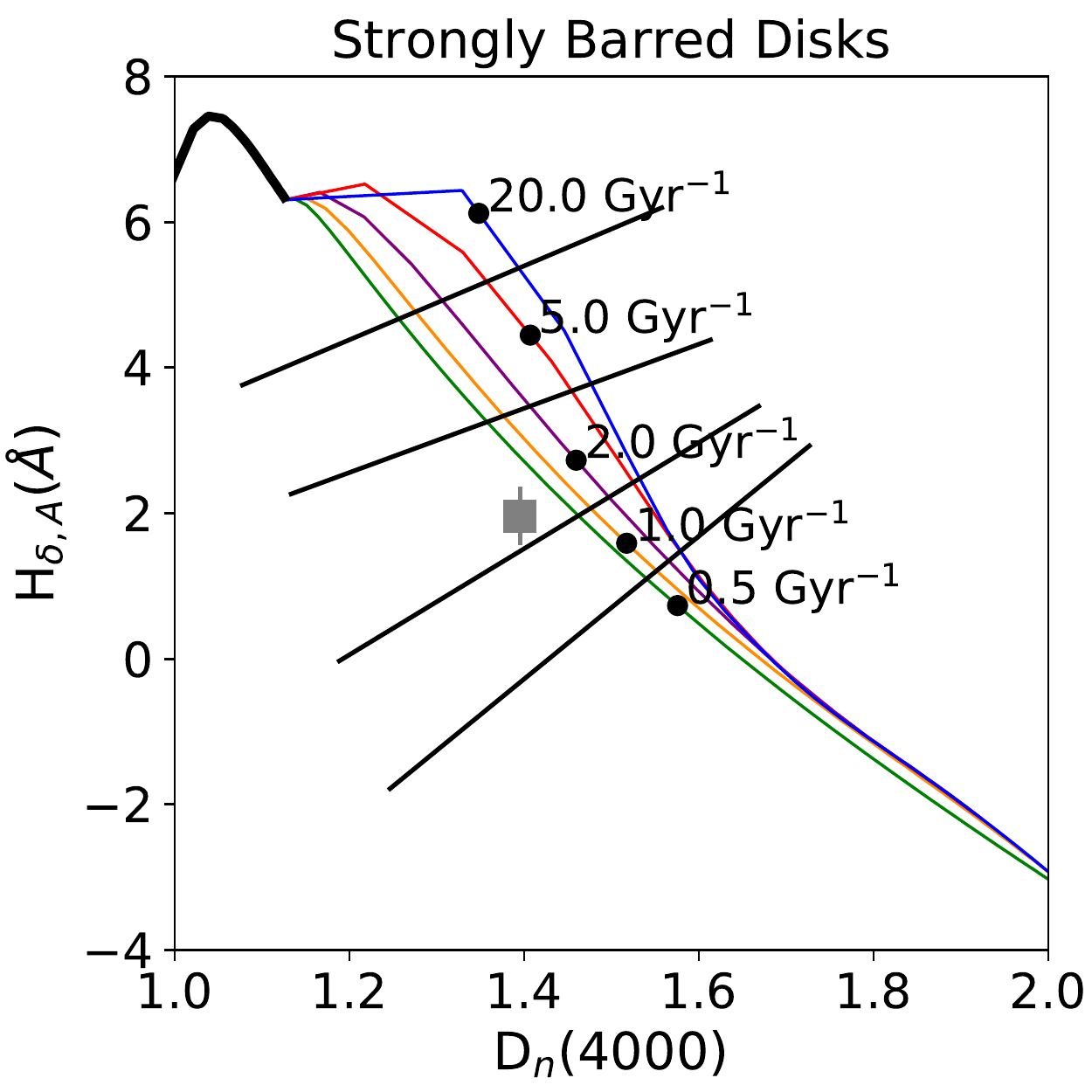}
\vspace*{0.5cm}
\includegraphics[width=5.5cm]{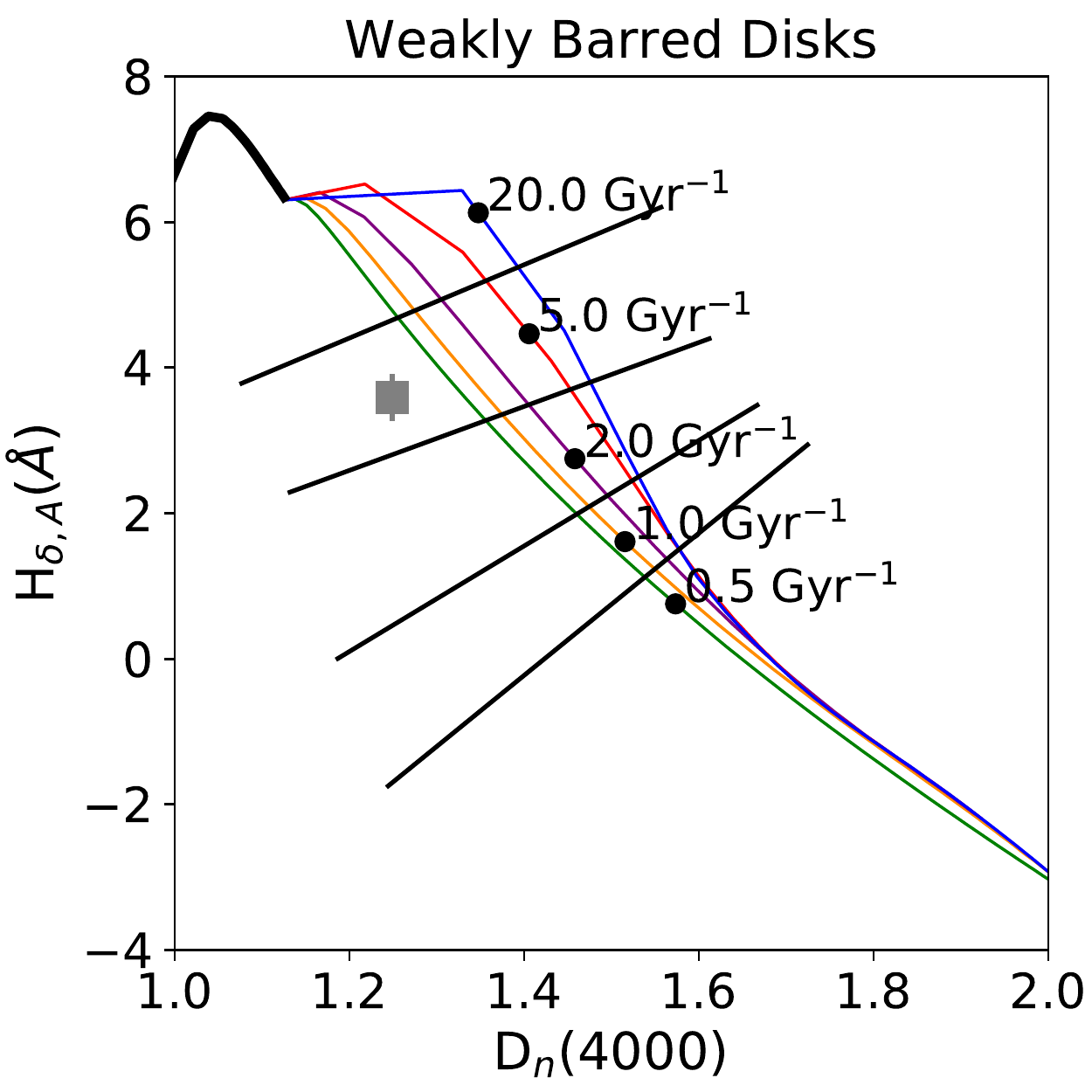}
\includegraphics[width=5.5cm]{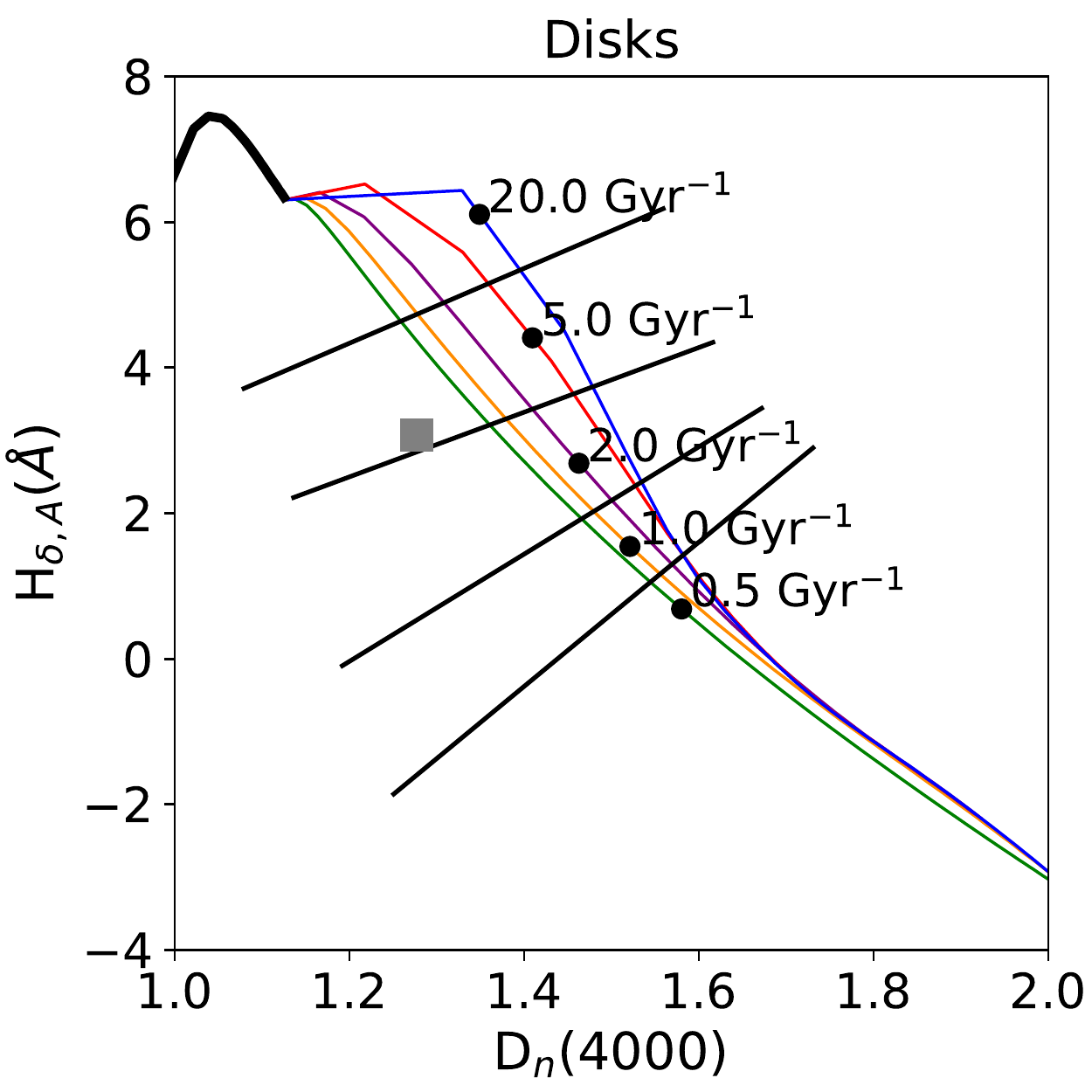}
\includegraphics[width=5.5cm]{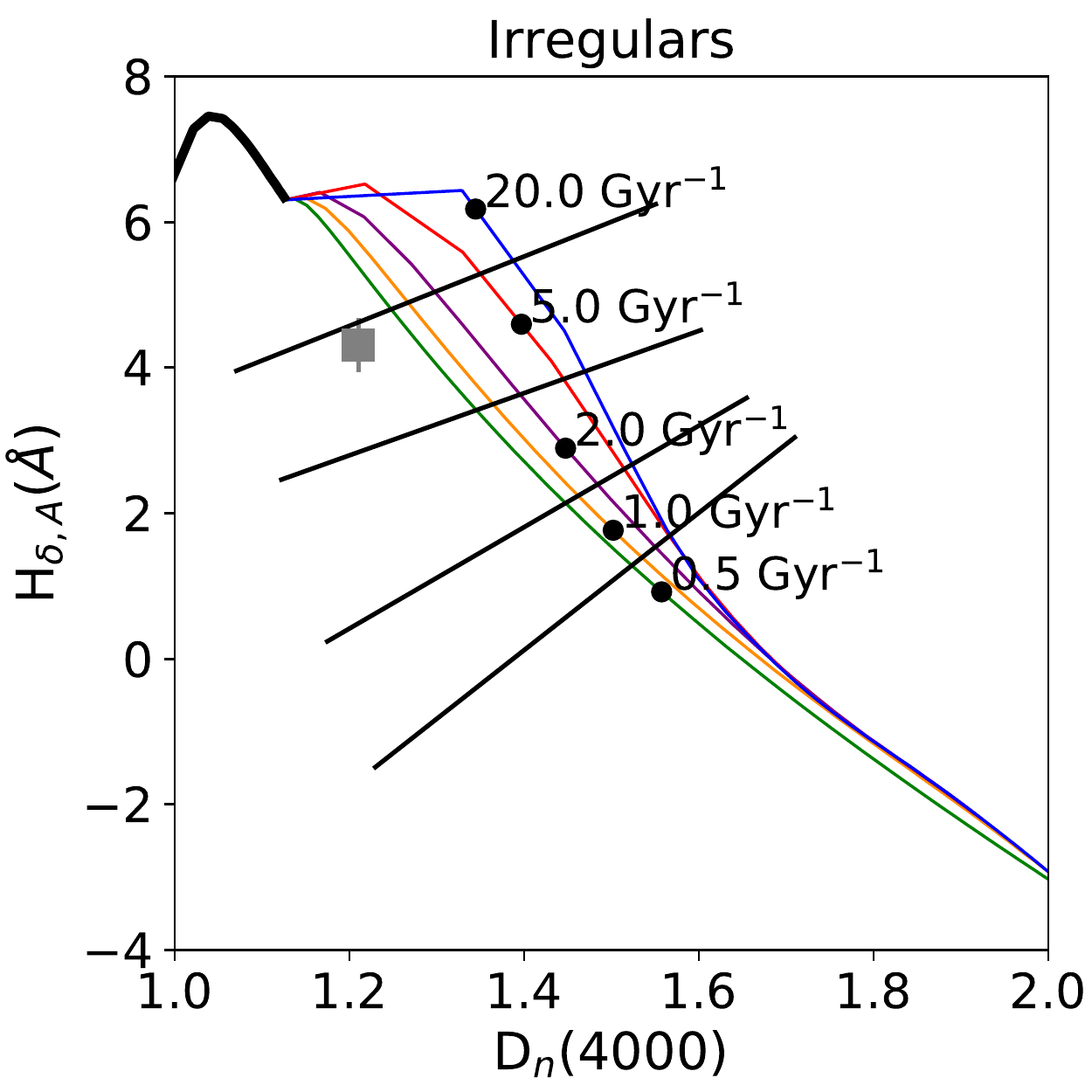}
\includegraphics[width=5.5cm]{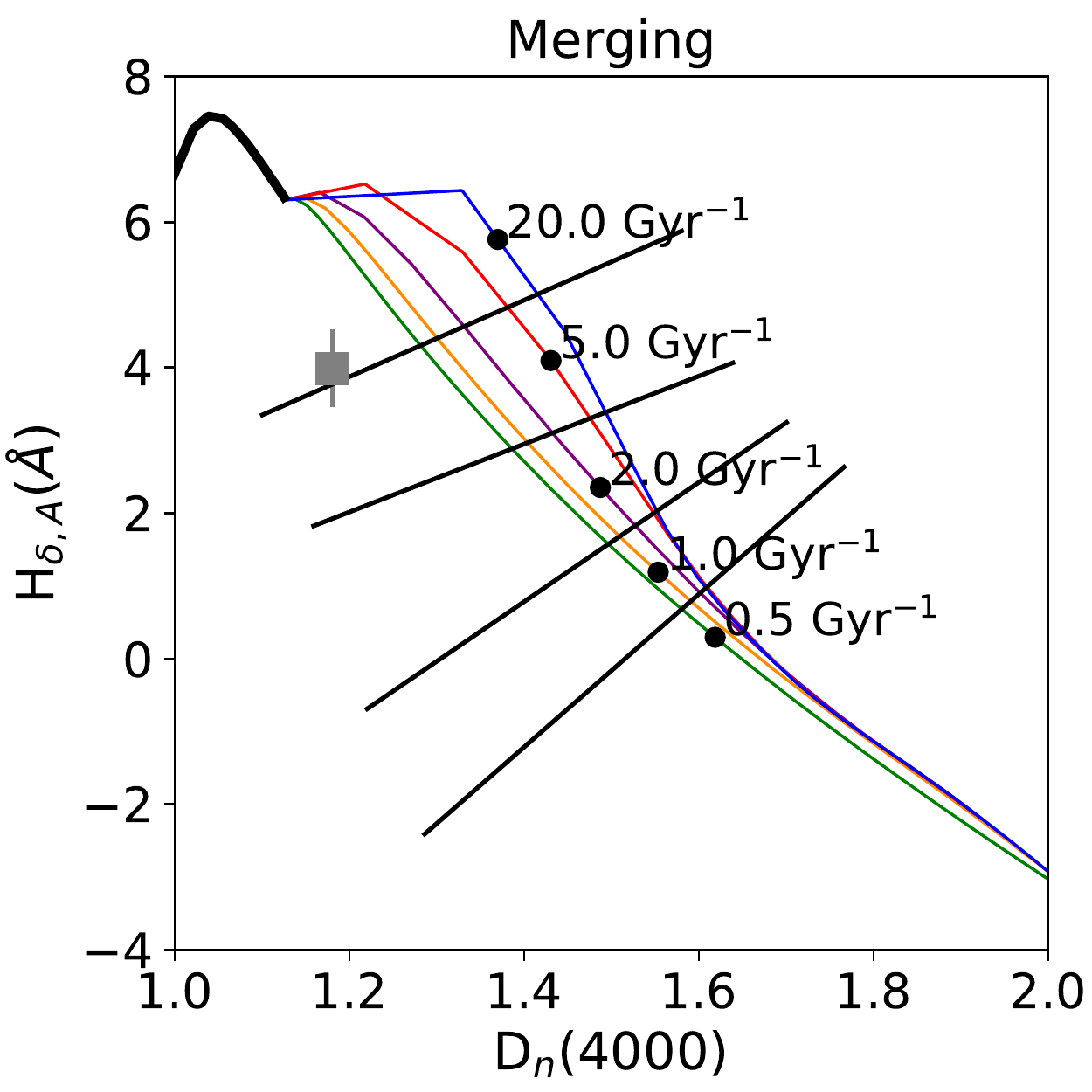}
\caption{H$_{\delta,A} \times$ D$_n(4000)$ plane and indices (grey squares) for each galaxy type. The black dots represent the D$_n(4000)$ and H$_{\delta,A}$ values in each SFH model when it reaches the mean galaxy colour specified on the upper right part in each plot. The error bars is the standard deviation of the indices. The straight lines represent the geometric mean in two consecutive SFH models.}
\label{gamma_values_cosmos_green_valley_galaxies}
\end{center}
\end{figure*}

\begin{figure*}
\begin{center}
\includegraphics[width=7.5cm]{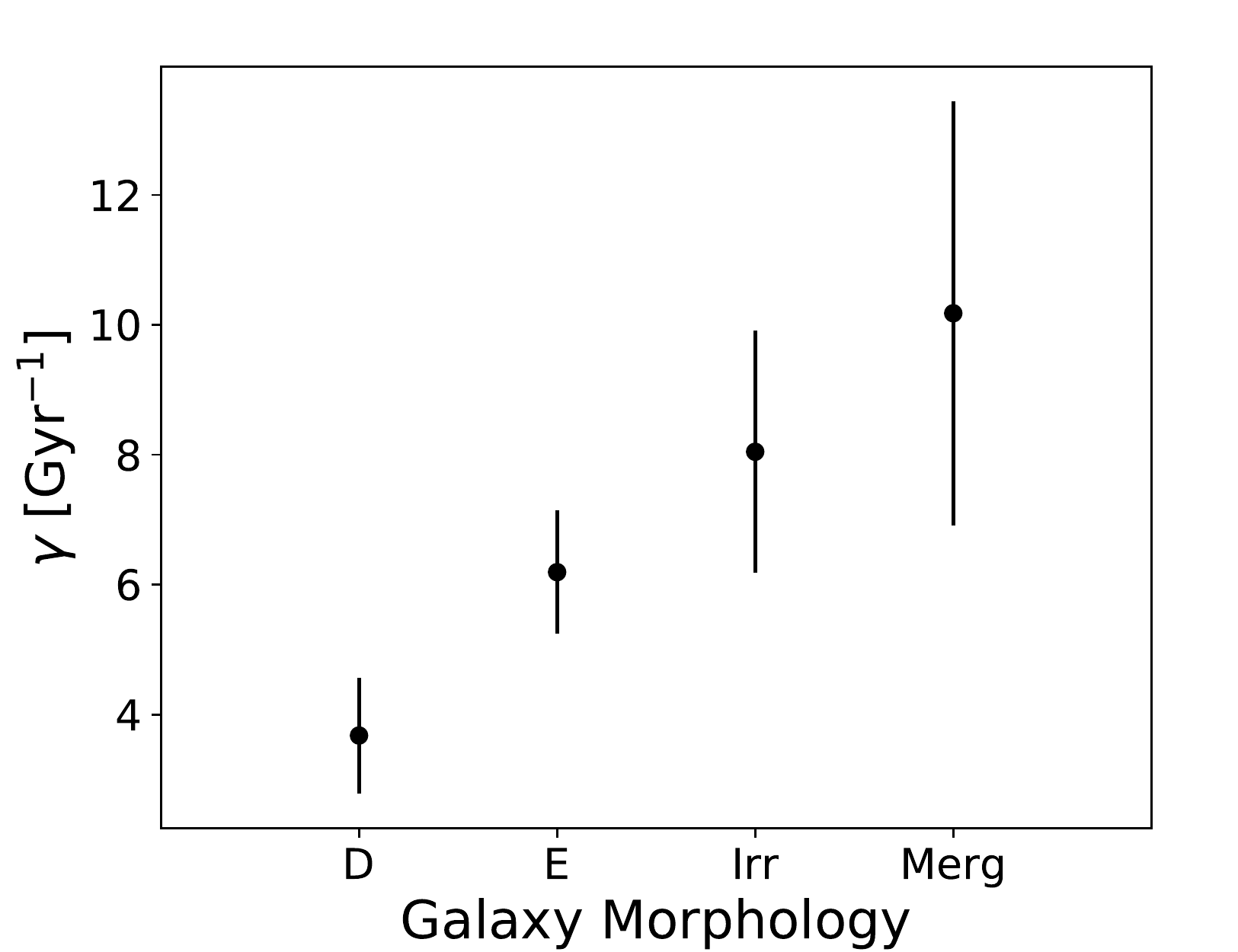}
\includegraphics[width=7.5cm]{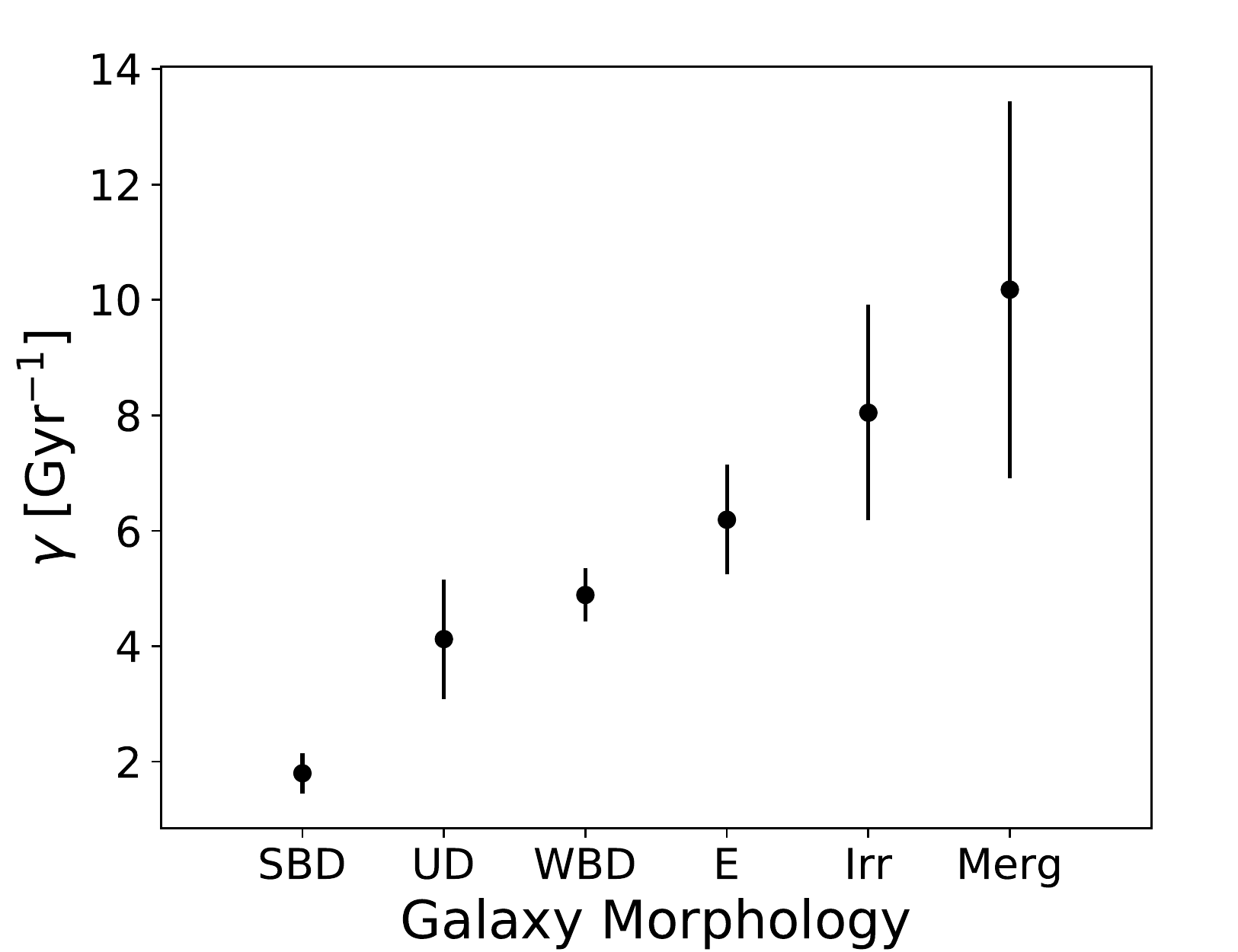}
\includegraphics[width=7.5cm]{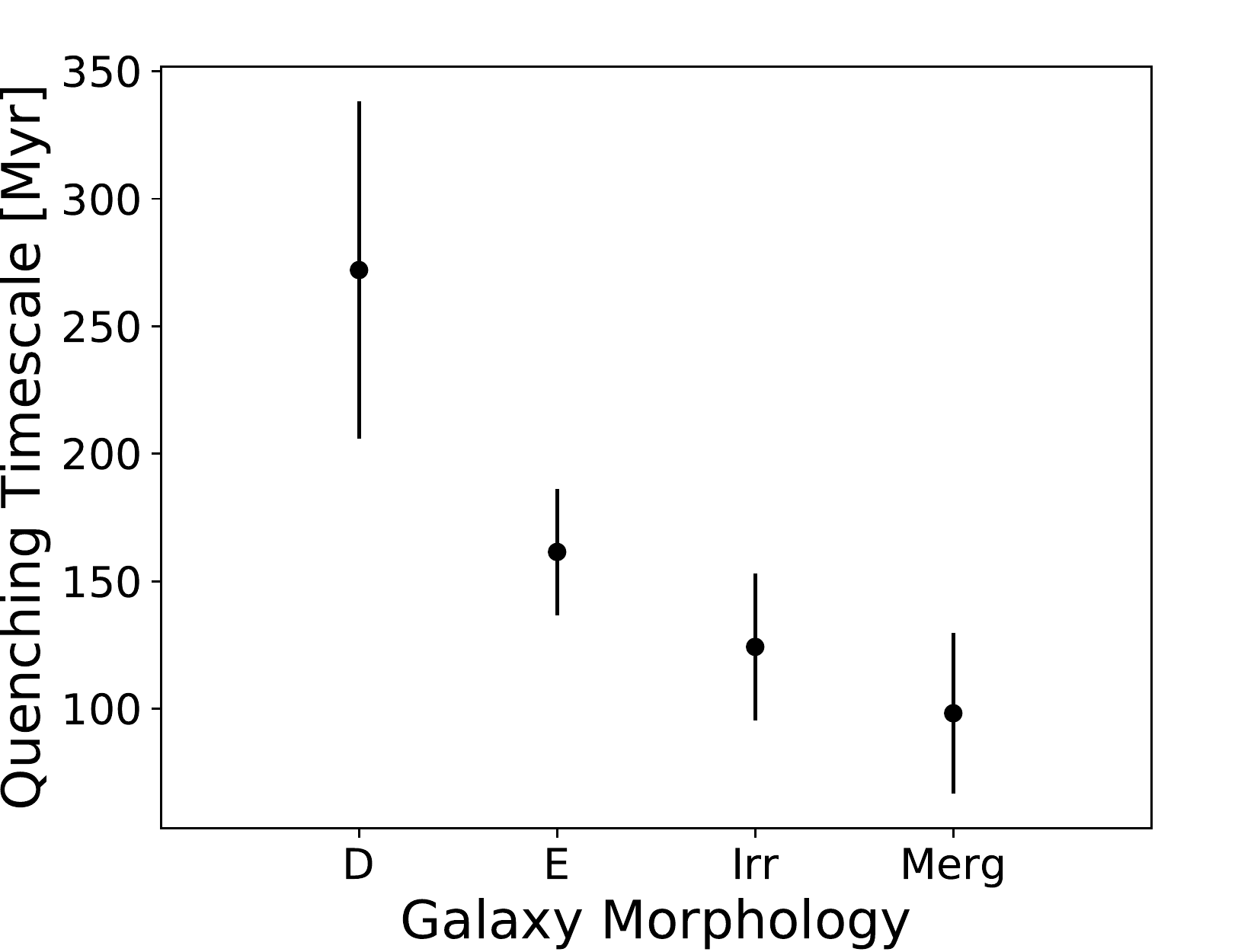}
\includegraphics[width=7.5cm]{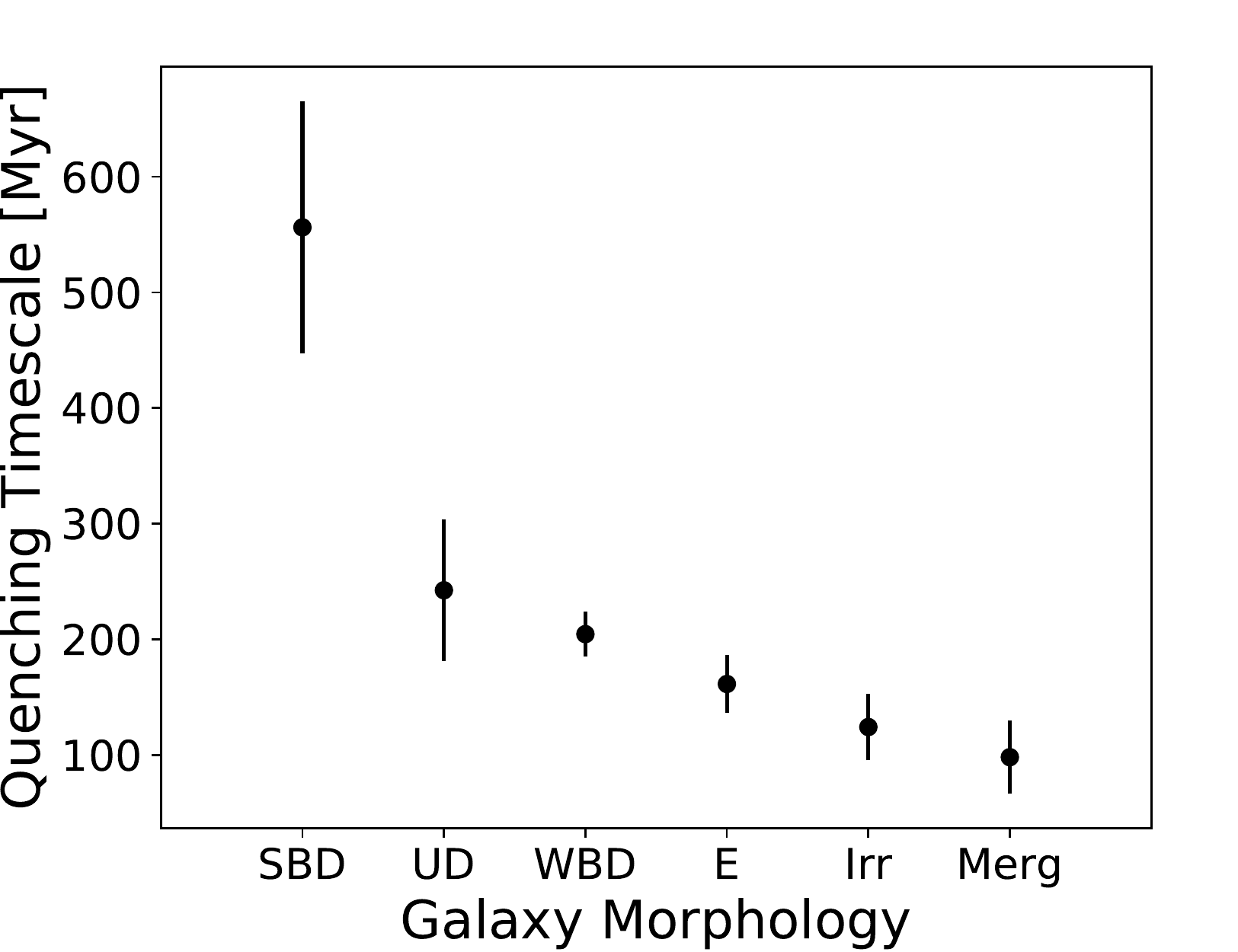}
\caption{\textit{\textbf{Top:}} Quenching index ($\gamma$) as a function of galaxy morphology. The left Figure shows the $\gamma$ values for the green valley galaxies classified as disk (D), ellipticals (E), irregulars (Irr) and merging galaxies (Merg). In the right Figure we divided the disk galaxies into strongly barred disks (SBD), unbarred disks (UD) and weakly barred disks (WBD). \textit{\textbf{Bottom:}} Quenching timescales when the initial star formation rate decreases to $\sim$37\% of the initial value ($1/\gamma$).}
\label{quenching_as_a_function_of_galaxy_morphology}
\end{center}
\end{figure*}

Figure \ref{gamma_values_cosmos_green_valley_galaxies} shows the position of the H$_{\delta,A}$ and D$_{n}$(4000) spectral indices measured for each of the coadded spectra shown in Figure \ref{coadded_zCOSMOS_spectra}. As mentioned in Section \ref{methodology}, in order to more adequately interpolate between the different SFH models, we show the geometric mean between two consecutive models with straight lines. The position of the measured H$_{\delta,A}$ and D$_{n}$(4000) with respect to these lines enables us to determine which SFH model is a better fit to the data and hence associate a star formation quenching index, $\gamma$ to the galaxy spectrum. For each morphological class considered in this work we repeat the measurement of $\gamma$ 1000 times, considering standard deviations in colour and spectral indices ($\left\langle \text{NUV}-r \right\rangle \pm \sigma_{\left\langle\text{NUV}-r\right\rangle}$, D$_{n}(4000)\pm \sigma_{\text{D}_{n}(4000)}$ and H$_{\delta,A}\pm \sigma_{\text{H}_{\delta,A}}$). Uncertainties in spectral indices are derived based on flux uncertainties of coadded spectra, shown as grey regions in Figure \ref{coadded_zCOSMOS_spectra}. Uncertainties in NUV$-r$ colours, on the other hand, are dominated by the standard deviation of colours of galaxies in each morphology bin (0.3-0.5 mag for different morphological types, compared to 0.15 mag for individual galaxies; see Section \ref{canada_france_hawaii_telescope_legacy_survey}). The final value of $\gamma$ and its corresponding uncertainty ($\sigma_{\gamma}$) are given by the average and the standard deviation, respectively, of the distribution of those measurements. These are shown in Table \ref{quenching_values} and Figure \ref{quenching_as_a_function_of_galaxy_morphology} for all morphological types considered in this work.

The distribution of these spectral indices from the green valley galaxy spectra on the D$_n$(4000) $\times$ H$_{\delta,A}$ plane approximately follows the curves predicted by our models, as demonstrated in early studies \citep{Kauffmann2003, Martin2007}. The left panel in Figure \ref{H_delta_A_vs_Dn4000_for_some_SDSS_GV_galaxies_with_SFH_models} shows the distribution of D$_n$(4000) and H$_{\delta,A}$ data points for randomly chosen green valley galaxies from the SDSS data at $z\sim0.2$. We note that although the assumed SFH model of a constant star formation rate followed by a period of exponential decay is somewhat simplistic, several studies support that an exponential decay in star formation activity is a good approximation at $z<1$ \citep[e.g.,][]{Baldry2008, Behroozi2013}. However, as a consequence of the complexity of galaxy SFHs, the measured spectral indices D$_n$(4000) and H$_{\delta,A}$ might deviate from the ones predicted by the models described in the Equations \ref{star_formation_history_1} and \ref{star_formation_history_2}. \citet{Martin2017} have shown, for example, that SFHs of green valley galaxies can be significantly more complicated than a simple quenching mechanisms, and a single galaxy could undergo several distinct quenching episodes.

As an example, \citet{Kauffmann2003} have shown that continuous SFHs lie systematically below the burst models, and we assume that any deviations of observed data from SFH models used in this work can be attributed to this simplification. For this reason we argue that absolute quenching timescales cited here should be taken with caution. Nevertheless, our main goal is to compare different quenching timescales for distinct galaxy populations, and in that sense our methodology is robust, since we are directly comparing average ages of the stellar population (as probed by the D$_n$(4000) index) and recent star formation episodes (as probed by the H$_{\delta}$ absorption from A-type stars) of galaxies with different morphologies.

Furthermore, even after taking these uncertainties into consideration, \citet{Martin2007} and \citet{Goncalves2012} have used the exact same methodology with great success, determining total mass fluxes across the green valley that agree well with the build-up of the stellar mass function in the red sequence since $z \sim 1$.

We find that disk galaxies have the longest star formation quenching timescales ($\gtrsim$250 Myrs), while green valley galaxies undergoing a merger quench their star formation $\sim$5 times faster, with a quenching timescale $\lesssim$50 Myrs. Within the disk classification, the strongly barred galaxies take the longest to have their star formation quenched, with a timescale close to $600$ Myrs. Weakly barred and unbarred galaxies transition through the green valley $\sim$3 times faster, with a star formation quenching timescale of $\sim$200 Myrs. Ellipticals and irregulars have on average quenching timescales below 200 Myrs.

%%%%%%%%%%%%%%%%%%
%% 
%% Figure 6
%% 
%%%%%%%%%%%%%%%%%%

%%%%%%%%%%%%%%%%%%
%% 
%% Figure 7
%% 
%%%%%%%%%%%%%%%%%%

%%%%%%%%%%%%%%%%%%
%% 
%% Table 1
%% 
%%%%%%%%%%%%%%%%%%

\begin{table*}
\begin{center}
\begin{tabular}{|c|c|c|c|}
\hline
Galaxy Type & Number of  & $\langle$NUV$-r\rangle$ & $\gamma$          \\
                        &  galaxies	  &    				            &  [Gyr$^{-1}$]        \\
\hline \hline
Unbarred Disks        & 94  & 3.5 $\pm$ 0.45 & 4.17 $\pm$ 0.37 \\
Strongly Barred Disks & 15  & 3.4 $\pm$ 0.35 & 1.7 $\pm$ 0.24 \\
Weakly Barred Disks   & 11  & 3.4 $\pm$ 0.23 & 4.79 $\pm$ 0.75 \\
Disks                 & 120 & 3.5 $\pm$ 0.42 & 3.66 $\pm$ 0.2 \\
Ellipticals           & 108 & 4.0 $\pm$ 0.35 & 6.08 $\pm$ 0.49 \\
Irregulars            & 48  & 3.4 $\pm$ 0.30 & 8.97 $\pm$ 1.73 \\
Merging               & 20  & 3.7 $\pm$ 0.50 & 20.43 $\pm$ 3.64 \\

\hline
\end{tabular}
\caption{Quenching indices and colour values for each galaxy type}
\label{quenching_values}
\end{center}
\end{table*}

%%%%%%%%%%%%%%%%%%
%% 
%% Discussion
%% 
%%%%%%%%%%%%%%%%%%

\section{Discussion}\label{discussion}

\begin{figure}
\begin{center}
\includegraphics[width = 7.5cm]{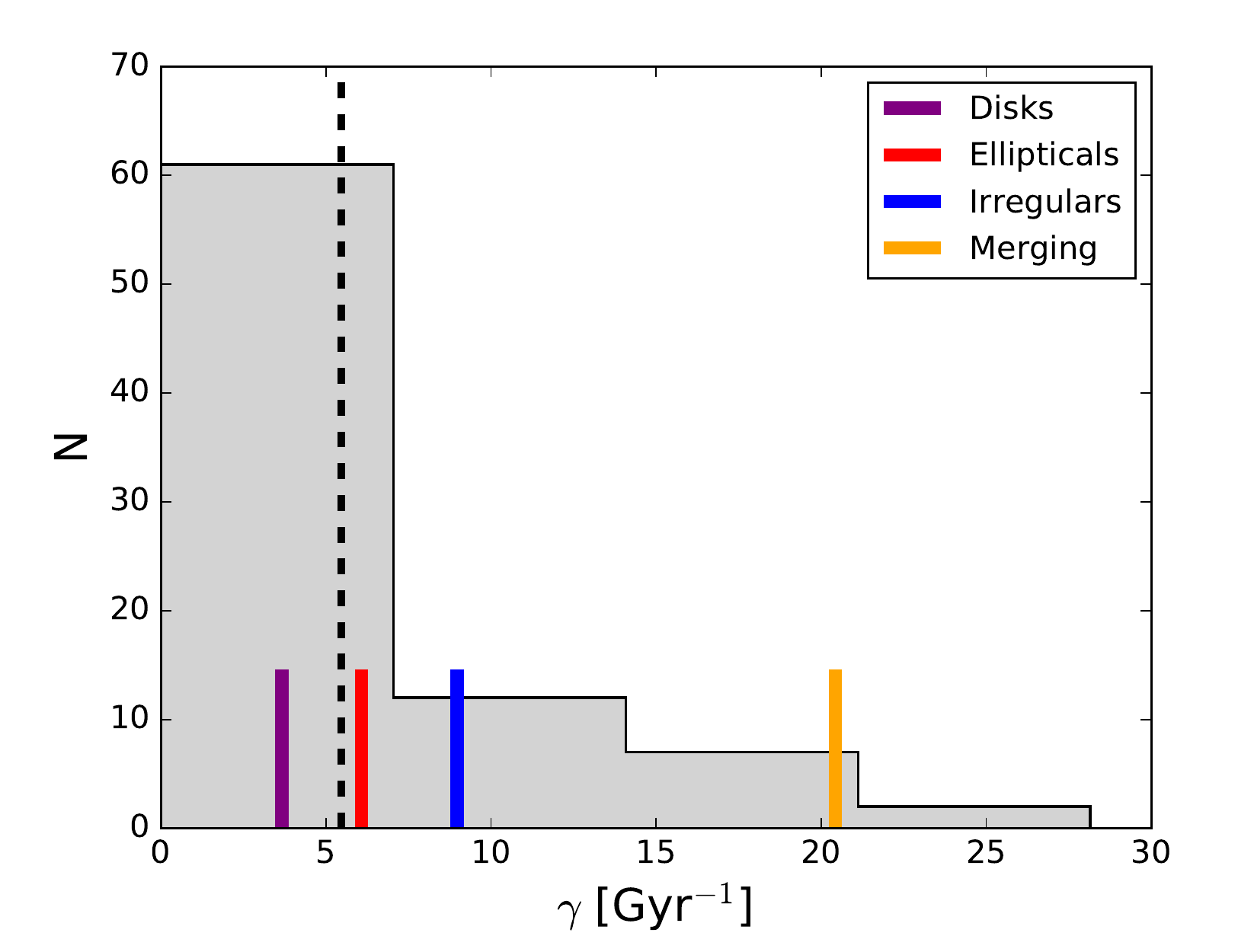}
\includegraphics[width = 7.5cm]{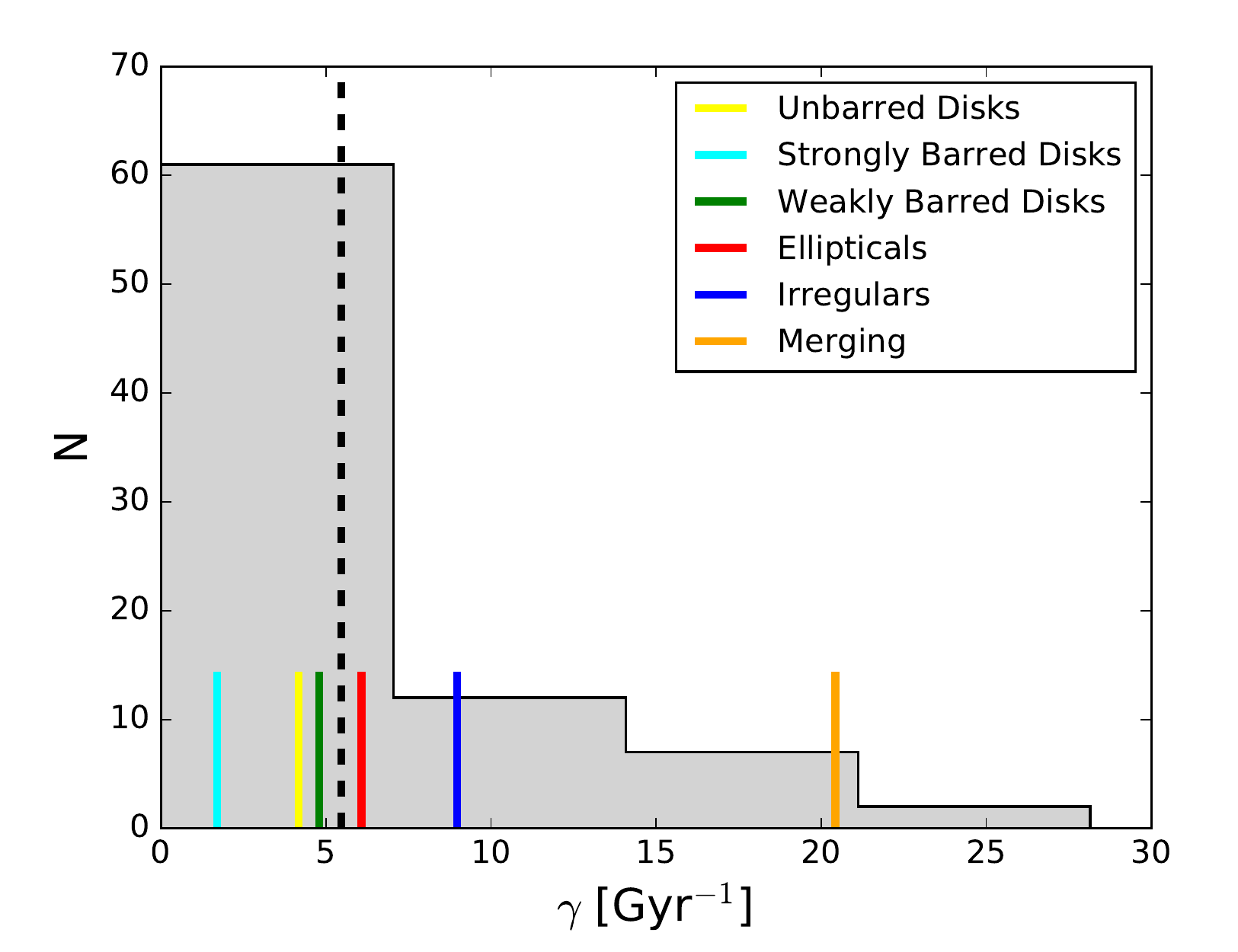}
\caption{Comparison between recalculated star formation quenching indices of \citet{Goncalves2012} and the star formation quenching indices of this work. The vertical dashed line represents the median value for the \citet{Goncalves2012} sample ($\gamma\sim 5$Gyr$^{-1}$) and the coloured full lines represent the average star formation quenching indices in this work. \textit{\textbf{Upper Panel:}} Comparison with the disk, elliptical, irregular and merging galaxies from this work. \textit{\textbf{Bottom Panel:}} Comparison with this work, considering the different disk subclasses (strongly barred, weakly barred and unbarred galaxies) independently. Taking the median value of \citet{Goncalves2012} as separating slow from fast quenching, the disk galaxies seem to be result of slower quenching, whereas elliptical, irregular and merging galaxies quench their star formation faster.}
\label{histograma_gammas}
\end{center}
\end{figure}

Figure \ref{histograma_gammas} shows our results, with star formation quenching timescales of $\gtrsim250$ Myrs for green valley disks and comparatively shorter values of $\sim150$ Myrs, $\sim100$ Myrs and $\sim50$ Myrs for ellipticals, irregulars and mergers, respectively. This indicates that intense processes $-$ with which the latter morphologies (ellipticals, irregulars and mergers) are typically associated $-$ are 60$\%$ to 5 times faster at winding down star formation activity in their host galaxies than disks.

We compare our derived star formation quenching timescales with those of \citet{Goncalves2012}, where the authors study a sample of $\gtrsim100$ green valley galaxies at $z\sim0.8$ based on deep individual optical spectra taken with the Keck DEIMOS spectrograph (see Figure \ref{histograma_gammas}). Although \citet{Goncalves2012} do not have morphological classifications for their galaxy sample, their results show star formation quenching timescales for a large sample of green valley galaxies within a similar redshift range as probed by our study. Considering that the bulk of our galaxies are classified as disks ($\sim$40\%) or ellipticals (43\%), our results are consistent with those found by \citet{Goncalves2012}, where the $\gamma$ distribution $-$ with a median value corresponding to $\sim$200 Myrs $-$ displays a clear trend towards the $\gamma$ values that we determined for these two morphological classes. Consistency of our coadd-based work with results based on individual galaxies demonstrates the robustness of our results: although there is likely diversity in star formation quenching timescales within any one morphological class, our coadded spectra reveal representative results as confirmed by comparing with an independent sample selection at similar redshifts.

%%%%%%%%%%%%%%%%%%
%% 
%% Figure 8
%% 
%%%%%%%%%%%%%%%%%%

Quenching timescales of green valley galaxies in our sample are generally shorter than those found in recent studies at lower redshifts. \citet{Martin2007} found that the typical quenching timescale in green valley galaxies at $z\sim0.1$ is $\gtrsim$ 1 Gyr (uncorrected for observability). This agrees with those values found for slowly evolving galaxies in the green valley as determined by \citet{Schawinski2014} and \citet{Smethurst2015} at $z<0.05$ ($\sim 1-2$ Gyr). These also agree with the hypothesis of starvation being the dominant quenching mode at later epochs \citep{Peng2015}, although a different parametrization of SFH in that work yields a timescale of 4--5 Gyr. Instead, our results show that galaxies at $z\sim0.8$ quench at timescales of a few hundred Myr, in excellent agreement with \citet{Goncalves2012} for the same epoch but based on a different dataset.

On one hand, we caution the reader to the fact that absolute values for quenching timescales are strongly dependent on the methodology, and should not be taken at face value, as discussed in Section \ref{introduction}. Nevertheless, the agreement with \citet{Goncalves2012} reinforces an interpretation of this evolution in quenching timescales as a downsizing, ``top-down'' scenario in which, at earlier times, massive galaxies are quenching faster, whereas in the local universe green valley galaxies are less massive and undergoing slower quenching mechanisms.

\citet{Schawinski2014} and \citet{Smethurst2015} also found that elliptical galaxies evolve more rapidly, with typical quenching timescales of approximately 250 Myr \citep{Schawinski2014}. Qualitatively, our results agree with these works, and our methodology confirms that elliptical galaxies quench faster than disks at all redshifts. On the other hand, our results indicate that even galaxies that are clearly undergoing secular evolution, such as strongly barred disks, present faster quenching timescales than similar objects at low redshift (up to 600 Myr, as opposed to a few Gyr). This could be the result of generally faster processes occurring for main-sequence galaxies at higher redshifts, as indicated for instance by shorter gas depletion times at earlier epochs \citep[700 Myr beyond $z\sim 1$,][]{Tacconi2013}.

%%%%%%%%%%%%%%%%%%
%% 
%% DISK/BAR (and fraction in GV) DISCUSSION
%% 
%%%%%%%%%%%%%%%%%%

The role of bars as major agents of disk secular evolution has been well established with various lines of observational evidence pointing to an enhancement in central molecular gas concentrations and an ensuing nuclear star formation activity \citep[e.g., ][]{Sheth2005, Ho1997, Kormendy2004}). Among our disk galaxy sub-samples, strongly barred galaxies display the longest star formation quenching timescales. We interpret this not as an indication that the presence of a bar delays in any way the star formation quenching process, but that the mere presence of a strong bar indicates that no other major quenching external agents have affected these galaxies as they transition through the green valley. These strongly-barred green valley galaxies thus represent sites of unencumbered transition from the blue cloud to the red sequence, without any external agent accelerating this process through star formation quenching $-$ neither major merger, nor cluster-related processes, such as harassment and ram-pressure stripping. On the other hand, unbarred and weakly-barred galaxies likely represent disks where external agents may be triggering a more rapid depletion of gas and hence a faster star formation quenching timescale, while interrupting the establishment of a strong bar. Furthermore, unbarred and weakly-barred galaxies may also have specific conditions (gas-richness and dynamical hotness) that on the one hand make them less hospitable towards (strong) bar formation \citep{Athanassoula2013}, but at the same time translate into a more gas-rich turbulent interstellar medium that is conducive to a faster depletion of the available molecular gas.

Systems corresponding to an evolutionary phase that is associated with long timescales are in principle expected to have a significant abundance in the sky. Disks make up almost half of the green valley population ($\sim43\%$) in our sample, consistent with the longer timescales we determined for their transition from the blue cloud to the red sequence. However, it may come as a surprise that barred galaxies only make up $\sim 9\%$ of the green valley sample, seemingly at odds with their long star formation quenching timescales. However, this scarcity of bars is a reflection of the low bar fraction at these redshifts, where bars make up $\sim22\%$ of the disk population, with a strong bar fraction of $\sim9\%$ \citep{Sheth2008}. A drop in the bar fraction has been attributed to a larger proportion of disks being dynamically-hot as we go to higher redshifts, hampering the establishment of bars \citep{Sheth2012}. Within our sample, the fraction of disks with a (strong) bar is ($13\%$) $22\%$, similar to that of the general disk population at these redshifts.Therefore, although the high fraction of disks that we find in our green valley sample is evidence of their longer quenching timescales, similarly-slow transitioning barred galaxies are simply too few in general at these redshifts to make up for a significant population in our sample.

%%%%%%%%%%%%%%%%%%
%% 
%% ELLIPTICALS DISCUSSION 
%% 
%%%%%%%%%%%%%%%%%%

The coadded spectra of galaxies displaying morphologies consistent with ellipticals, irregulars and mergers point to faster star formation quenching timescales. We note that visual inspection of HST images for our irregular sample (see Figure \ref{green_valley_galaxy_images}) reveals that these galaxies are more akin clumpy ones at higher redshits. These systems have been found to be quite abundant at $z\sim1-2$ \citep{Bournaud2015}. The stellar mass of such galaxies is in the range of $10^{10}-10^{11}$ M$_{\odot}$ and they have been attributed with short gas consumption timescales \citep{Bournaud2015}, consistent with the short star formation quenching timescales we derive (see Figure \ref{quenching_as_a_function_of_galaxy_morphology}). \citet{Bournaud2015} discuss that clumpy galaxies have very irregular morphologies in the optical when compared to nearby spirals of similar mass, because the high gas fractions cause strong gravitational instabilities in the galactic disks, resulting in disk fragmentation and clump formation, with clump masses in the order of $10^8-10^9$ M$_{\odot}$. These giant clumps may be at the origin of the formation of bulges in galaxies. However, \citet{Puech2010} argues that clumpy galaxies at $z<1$ are more stable than those at higher redshifts, suggesting a scenario where interactions between galaxies could be the dominant driver for the formation of clumpy galaxies at $z<1$. Therefore, they could be coalesced mergers (i.e., mergers in a more advanced stage). Regardless of the clumpy origin, it is expected that these galaxies have very rapid star formation quenching timescales.

The fraction of elliptical galaxies in our green valley sample ($\sim$40\%) is almost as high as the fraction of disk galaxies (43\%), suggesting that the physical processes responsible for elliptical galaxy formation are common. It has been widely established that the characteristically old stellar population associated to the bulk of the stellar content in massive ellipticals formed in a single event in the distant past, likely a major merger \citep{Toomre1977}. Merger simulations indicate that the interaction between colliding gas-rich spirals may extend over $\sim1-2$ Gyr \citep[e.g., ][]{Springel2005d, Conselice2006}, with a sudden burst of star formation lasting $\sim$10$^8$~yr. Together with the suggestion by \citet{Peng2010} that galaxy mergers are one of the main mechanisms to quench star formation at $z>0.5$, the abundance of this morphological class may be explained with a galaxy merger origin. Applying an exponential decay model for star formation rates on SDSS photometric data, \citet{Schawinski2014} reached a similar conclusion where elliptical galaxies at low redshifts quench their star formation activity through mergers. 

We find increasingly short timescales for the elliptical, irregular and merger morphological classifications and interpret them as reflecting three different time perspectives of a sequence of a major galaxy fusion. We attribute this to the fact that we are measuring the \textit{average} star formation quenching timescales for each galaxy type from the galaxy spectra and photometry. The set of galaxies composing our ``merger" sample corresponds to systems caught early-on in a merger sequence, when interacting members are still distinguishable from each other. Rest-frame NUV-r colours place these systems within the green valley and the short quenching timescales are a reflection of a short-lived peak and drop in star-formation that triggers a rapid change in the system's colour. Irregulars may in turn correspond to a more advanced merger phase, when the interacting galaxies have coalesced to a single, disturbed system that displays multiple knots likely corresponding to the merging galactic nuclei. The average star formation quenching timescale hence does not merely include a sharp change in colour, but a phase in which the coalesced system likely does not alter significantly its global colour. Lastly, the galaxies with spheroidal morphology $-$ classified as ellipticals $-$ denote the end-point stage of a post-merger system with no remaining morphological signature of a past interaction. Considering that the rest-frame, dust-corrected colours of these galaxies place them within the green valley region of the CMD, we are catching them at a transitional stage. While galaxies caught at an early merger stage are in the midsts of a very sharp change in global colour, irregulars and $-$ more significantly $-$ ellipticals trace a more diluted change in colour due to a merger in their recent past. The proportion of these galaxy types $-$ mergers ($7\%$), irregulars ($17\%$) and ellipticals ($40\%$) $-$ within our green valley sample are consistent with a picture where these three morphological classifications trace three distinct stages in a merger sequence, with increasingly longer lifetimes. With an increase rate in merger activity \citep{Lotz2011}, the abundance of clumpy galaxies and merger-like morphologies in the green valley is expected to increase with redshift.

Based on the analysis of SDSS photometry for green valley galaxies at $z\sim0.2$, \citet{Schawinski2014} suggested that secular evolution and mergers as the two main scenarios for star formation quenching at low redshift. With more than $80\%$ of our green valley sample represented by ellipticals and disks in similar proportions ($\sim40$\% and $\sim43$\%, respectively), our results at intermediate redshifts ($z\sim0.5-1$) are consistent with both of these agents being important drivers of the galaxy colour transition at higher redshifts, at least at $z<1$. However, our quantification of the star formation quenching timescale as a function of morphology indicates that although secular evolution may ultimately lead to gas exhaustion in the host galaxy via bar-induced gas inflows that trigger star formation activity, secular agents are not major agents in the rapid quenching of galaxies at these redshifts. Galaxy interaction, associated with the elliptical, irregular and merger morphologies contribute, to a more significant degree, to the fast transition through the green valley at these redshifts. Our results provide an explanation to the recent findings that star formation quenching happened at a faster pace at $z\sim0.8$ \citep{Goncalves2012}.

%%%%%%%%%%%%%%%%%%
%% 
%% Summary
%% 
%%%%%%%%%%%%%%%%%%

\section{Summary}\label{summary}

Average gas consumption timescales for main-sequence star-forming galaxies of $\sim 2$ Gyrs in the local universe \citep{Bigiel2008} and  $\sim700$ Myrs at $z=1-3$ \citep{Tacconi2013} are not consistent with the scarcity of galaxies within the green valley region of the colour-magnitude diagram and the observed growth of the red sequence of galaxies. With these timescales, the natural gas consumption caused by star formation as observed in the main-sequence of star-forming galaxies should in principle lead galaxies to undergo a relatively smooth colour transition through the green valley. Therefore, it becomes clear that other astrophysical agents must be responsible for triggering sudden episodes of intense gas exhaustion (e.g., nuclear and/or galaxy-wide starbursts) that rapidly drive the galaxy-wide colour transition at a faster pace. Indeed, past studies have shown the timescale for star formation quenching to be closer to $\sim1$ Gyr in the local universe \citep{Martin2007} and $\sim200$ Myrs in $z\sim0.8$ \citep{Goncalves2012}.

In order to elucidate the role of the nature of the processes that drive the rapid colour transition through the green valley, we measured star formation quenching timescales for different galaxy morphological types, associating different morphologies with physical processes that may not only change the galaxy morphological structure but also stop star formation activity. Based on  the photometric data of the Canada-France-Hawaii Telescope Legacy Survey (CFHTLS) we identify our green valley parent sample. We cross-matched the CFHTLS green valley galaxies with: (1) the 10k zCOSMOS spectroscopic survey; (2) the strongly- and weakly-barred galaxy catalog by \citet{Sheth2008}; and (3) the Zurich Estimator of Structure Types Catalog, which contains information about galaxy morphologies. In addition, we added a ``merger" galaxy type based on visual inspection. We calculate star formation quenching timescales following the approach adopted by \citet{Martin2007} and \citet{Goncalves2012}, based on the galaxy NUV$-r$ colour, measuring the H$_{\delta,A}$ and D$_{n}(4000)$ spectral indices,  and assuming an exponential decay of the galaxy star formation history. Our main results are:

\begin{enumerate}

\item With $>80\%$ of our green valley sample displaying disk-like or spheroidal morphologies consistent with ellipticals, our work suggests that the physical agents associated with these morphologies are common.

\item We derive star formation quenching timescales for green valley disks that are up to 5 times longer than those found for morphologies associated to merging activity. 

\item The star formation quenching timescale in barred galaxies is the lowest among all galaxy types, suggesting that these are sites where no other major quenching external agents have had an impact as they transition through the green valley.

\item Although barred galaxies only make up $\sim 9\%$ of the green valley sample, seemingly at odds with their long star formation quenching timescales, we attribute the scarcity of bars in the green valley to the low bar fraction at these redshifts.Within our sample, the fraction of barred disks is $22\%$, similar to that of the general disk population at these redshifts \citep{Sheth2008}.

\item We find increasingly short timescales for the elliptical, irregular and merger morphologies. Considering that we measure the \textit{average} star formation quenching timescales, it becomes clear that while galaxies caught at an early merger stage are in the midsts of a very sharp change in global colour, irregulars (which we associate to coalesced mergers) and $-$ more significantly $-$ ellipticals trace a more diluted change in colour due to a merger in their recent past, followed by a phase in which the galaxy does not significantly alter its global colour. 

\end{enumerate}

Our analysis of star formation quenching timescales as a function of galaxy morphology at intermediate redshifts ($z\sim0.5-1$) suggests that both secular evolution and merger activity are common processes within the green valley. However, although secular evolution does appear to play a role in the gas exhaustion in green valley galaxies, the longer timescales associated to barred galaxies relative to those of morphologies reminiscent of merger activity suggests that they are not the main driver of the rapid quenching of galaxies at these redshifts. Galaxy interaction, associated with the elliptical, irregular and merger morphologies contribute, to a more significant degree, to the fast transition through the green valley at these redshifts. This in turn provides a partial explanation as to why star formation quenching happens at a faster pace at $z\sim0.8$ \citep{Goncalves2012}. With an increase in merger activity expected at higher redshifts, green valley galaxies appear to be undergoing global colour transformation due to a higher incidence of mergers that in turn lead to a faster migration through the green valley in the colour magnitude diagram of the general galaxy population. Still, our measurements indicate that even disk galaxies quench at a fast pace at earlier times, leading to the conclusion that the downsizing of quenching timescales since $z\sim 1$ occurs for all morphological types.

\section*{Acknowledgements}

JPNC was supported by a PhD grant from CAPES $-$ the Brazilian Federal Agency for Support and Evaluation of Graduate Education within the Ministry of Education of Brazil. TSG and KMD thank the support of the Productivity in Research Grant of the Brazilian National Council for Scientific and Technological Development (CNPq). KS acknowledges support from the National Radio Astronomy Observatory, a facility of the National Science Foundation operated under cooperative agreement by Associated Universities, Inc. This work was conducted during a scholarship held by KMD and supported by the Brazilian Science Without Borders program, managed by CAPES and CNPq . 

We thank the Laborat\'orio de Astrof\'isica Extragal\'actica (LASEX) \footnote{http://dgp.cnpq.br/dgp/espelhogrupo/5167044310442074} from Observat\'orio do Valongo for fruitful discussions, in particular Laurie Riguccini and Ald\'ee Charbonnier. 

%%%%%%%%%%%%%%%%%%%%%%%%%%%%%%%%%%%%%%%%%%%%%%%%%%



%\bibliographystyle{mnras}


\bibliographystyle{mnras}
\bibliography{Referencias_bibliograficas_extensao_bib/Ref_compact_galaxies,Referencias_bibliograficas_extensao_bib/Ref_environment,Referencias_bibliograficas_extensao_bib/Ref_galaxy_bimodality,Referencias_bibliograficas_extensao_bib/Ref_galaxy_zoo,Referencias_bibliograficas_extensao_bib/Ref_GALEX,Referencias_bibliograficas_extensao_bib/Ref_morphological_classification,Referencias_bibliograficas_extensao_bib/Ref_green_valley,Referencias_bibliograficas_extensao_bib/Ref_passive_gals_are_more_compacts_than_normal_ones,Referencias_bibliograficas_extensao_bib/Ref_red_sequence_growth,Referencias_bibliograficas_extensao_bib/Ref_SDSS,Referencias_bibliograficas_extensao_bib/Ref_star_forming_main_sequence,Referencias_bibliograficas_extensao_bib/Ref_secular_evolution,Referencias_bibliograficas_extensao_bib/Ref_galaxy_mergers,Referencias_bibliograficas_extensao_bib/Ref_AGN,Referencias_bibliograficas_extensao_bib/Ref_stellar_feedback,Referencias_bibliograficas_extensao_bib/Ref_star_formation_history_and_stellar_populations,Referencias_bibliograficas_extensao_bib/Ref_galaxy_stellar_mass_function,Referencias_bibliograficas_extensao_bib/Ref_k_correction,Referencias_bibliograficas_extensao_bib/Ref_dust_in_galaxies,Referencias_bibliograficas_extensao_bib/Ref_clumpy_galaxies,Referencias_bibliograficas_extensao_bib/Ref_correlation_central_black_hole_with_host_galaxy,Referencias_bibliograficas_extensao_bib/Ref_morphological_quenching,Referencias_bibliograficas_extensao_bib/Ref_gas_content_and_gas_depletion_over_cosmic_times,Referencias_bibliograficas_extensao_bib/Ref_downsizing,Referencias_bibliograficas_extensao_bib/Ref_COSMOS}

%%%%%%%%%%%%%%%%%%%%%%%%%%%%%%%%%%%%%%%%%%%%%%%%%%

%%%%%%%%%%%%%%%%% APPENDICES %%%%%%%%%%%%%%%%%%%%%

% \appendix
% 
% \section{Some extra material}
% 
% If you want to present additional material which would interrupt the flow of the main paper,
% it can be placed in an Appendix which appears after the list of references.

%%%%%%%%%%%%%%%%%%%%%%%%%%%%%%%%%%%%%%%%%%%%%%%%%%

% Don't change these lines
\bsp	% typesetting comment
\label{lastpage}
\end{document}